\documentclass[reprint,twocolumn,superscriptaddress,showpacs,preprintnumbers,nofootinbib,aps,prd]{revtex4-1}

\normalsize

\usepackage{graphicx,color}
\usepackage{amsmath,amssymb,amsfonts}
\usepackage{stackrel}
\usepackage{enumitem}
\usepackage[english]{babel}
\usepackage{verbatim}
\usepackage{tabularx,booktabs}
\usepackage{comment}
\usepackage[normalem]{ulem}
\usepackage{xcolor}
\usepackage{subcaption}
\usepackage{dcolumn}
\usepackage{physics}
\usepackage{bbm}
\usepackage{setspace}
\usepackage{float}

\usepackage{hyperref}
\hypersetup{
    pdfnewwindow=true,   
    colorlinks=true,     
    linkcolor=blue,      
    citecolor=purple,      
    filecolor=blue,      
    urlcolor=blue        
}
 
\graphicspath{
  {./}
  {figures/}
}

\renewcommand{\lq}{``}
\renewcommand{\rq}{''}

\definecolor{dark-green}{rgb}{0.1,0.7,0.3}

\newcommand{\tf}{\texorpdfstring}

\newcommand{\tev}{~\text{TeV}}
\newcommand{\mev}{~\text{MeV}}

\newcommand{\onbb}{0\nu\beta\beta}

\def\nn{\nonumber}

\newcommand{\UMASS}{\affiliation{Amherst Center for Fundamental Interactions, Department of Physics, University of Massachusetts, Amherst, MA 01003, USA}}
\newcommand{\TDLI}{\affiliation{Tsung-Dao Lee Institute and School of Physics and Astronomy, Shanghai Jiao Tong University, 800 Dongchuan Road, Shanghai, 200240 China}}

\newcommand{\CALTECH}{\affiliation{Kellogg Radiation Laboratory, California Institute of Technology, Pasadena, CA 91125, USA}}
\newcommand{\SUNYATSET}{\affiliation{School of Physics and Astronomy, Sun Yat-sen University, Zhuhai 519082, P.R. China.}}
\newcommand{\INT}{\affiliation{Institute for Nuclear Theory, University of Washington, Seattle, WA 98195-1550, USA}}
\newcommand{\NU}{\affiliation{Northwestern University, Department of Physics \& Astronomy, 2145 Sheridan Road, Evanston, IL 60208, USA}}

\begin{document}
\title{ Testable leptogenesis and \texorpdfstring{$0\nu\beta\beta$}{0nubb} decay in extended seesaw model}
\author{Gang Li}
\email{ligang65@mail.sysu.edu.cn}
\SUNYATSET

\author{Michael J. Ramsey-Musolf}
\email{mjrm@sjtu.edu.cn, mjrm@physics.umass.edu}
\TDLI
\UMASS
\CALTECH

\author{Supriya Senapati}
\email{ssenapati@umass.edu}
\UMASS

\author{Sebasti\'an Urrutia Quiroga}
\email{suq90@uw.edu}
\UMASS
\INT
\NU

\begin{abstract}
\vspace*{0.5cm}

We investigate the possibility of observable neutrinoless double beta decay $( 0 \nu \beta\beta)$ and viable leptogenesis within a low-scale extended inverse seesaw mechanism with additional sterile neutrinos. General effective field theory (EFT) considerations suggest that if there are experimentally observable signatures in $0 \nu \beta \beta$-decay and a lepton asymmetry generated by heavy right-handed neutrino decays, thermal leptogenesis is likely to be unviable. However, in this work, we show that in the context of low-scale leptogenesis, one can obtain the observed baryon asymmetry of the universe and observable signatures of $0 \nu \beta \beta$ decay in the presence of additional sterile neutrinos. In this framework, the light neutrino masses are suppressed by the extended seesaw parameter, $\mu$, thereby allowing for $\mathcal{O}(10\, \mathrm{TeV})$ right-handed (RH) neutrinos, while avoiding small Yukawa couplings as in other neutrino mass models and near degeneracies in the RH neutrino spectrum as required by the low-scale leptogenesis paradigm. Contributions to the $0 \nu \beta \beta$-decay rate from additional sterile neutrinos can be appreciable, while the corresponding contributions to the early universe lepton asymmetry washout rate are suppressed by other parameters not entering the $0 \nu \beta \beta$-decay amplitudes. We show that for keV-MeV scale, sterile neutrinos 
future ton-scale $0\nu\beta\beta$-decay experiments offer potential signals while maintaining viable leptogenesis.

\preprint{ACFI-T24-07, INT-PUB-24-051}

\end{abstract}

\maketitle

\section{Introduction}
\label{sec:intro}

In the Standard Model (SM) of particle physics, lepton number remains conserved at the classical (Lagrangian) level. The well-known seesaw mechanism~\cite{Minkowski:1977sc,GellMann:1980vs,Yanagida:1979as,Mohapatra:1979ia,Cheng:1980qt,Lazarides:1980nt,Magg:1980ut,Schechter:1980gr,Foot:1988aq,Ma:1998dn}, explaining the origin of small neutrino masses, strongly suggests the presence of overall lepton number violating (LNV) interactions in extensions to the SM. The minimal approach to account for massive neutrinos embedded into the SM framework is
the seesaw mechanism (types I, II, and III). However, in these scenarios, the scale of the beyond SM particles (right-handed neutrinos, scalar triplet, or fermionic triplet) is required to be very high ($\gtrsim 10^{14}$ GeV) to generate correct light neutrino masses; in turn, these heavier states cannot be experimentally observed. 

As an alternative, one can generate neutrino mass by low-scale seesaw mechanisms like the linear seesaw~\cite{Wyler:1982dd,Akhmedov:1995vm, Akhmedov:1995ip}, inverse or double seesaw~\cite{Wyler:1982dd,Mohapatra:1986bd,Barr:2003nn,Deppisch:2004fa}.
However, these mechanisms require additional sterile neutrinos with the SM particle content with masses around the electroweak scale or even lower. Hence, the new states can be produced directly in colliders and generate observable signatures in low-energy experiments such as neutrinoless double beta ($0\nu\beta\beta$) decay.

Simultaneously, LNV interactions can play a crucial role in generating the baryon asymmetry of the universe (BAU), typically quantified in terms of the baryon-to-photon number density based on PLANCK 2018 data~\cite{ParticleDataGroup:2020ssz, Planck:2018vyg}:
$ \eta_B^{\text{obs}} = {n_B}/{n_\gamma} = (6.12 \pm 0.04) \times 10^{-10} $. This value aligns with constraints from Big Bang Nucleosynthesis~\cite{ParticleDataGroup:2020ssz}, serving as a key test for standard cosmology. Given that the SM fails to satisfy the Sakharov conditions~\cite{Sakharov:1967dj} for explaining the observed baryon asymmetry, physics beyond the Standard Model (BSM) is required. One possibility, baryogenesis via leptogenesis, entails LNV interactions generating a sufficient amount of asymmetry within the lepton sector via CP-violating, out-of-equilibrium decays of the new seesaw mechanism states ({\it e.g.,} right-handed Majorana neutrinos). 
This asymmetry is subsequently converted into the observed baryon asymmetry through non-perturbative sphaleron processes before the electroweak phase transition~\cite{Fukugita:1986hr}. Among various proposed mechanisms for the dynamical baryon asymmetry generation, leptogenesis attracts attention as it connects the baryon asymmetry with the small neutrino masses. Thus, observing LNV interactions could have profound implications, providing clues not only about a Majorana contribution to neutrino masses but also influencing the validity of various leptogenesis scenarios.

In this context, the mass scale $\Lambda$ associated with the BSM lepton number violation is a decisive factor. For $\Lambda$ well below the conventional thermal leptogenesis scale, characterized by the Davidson-Ibarra lower bound~\cite{Davidson:2002qv}, viable leptogenesis necessitates alternative scenarios, such as resonant~\cite{Blanchet:2006be, Deppisch:2010fr} and ARS~\cite{Akhmedov:1998qx, Asaka:2005pn, Canetti:2012kh, Chun:2017spz, Abada:2018oly, Das:2024gua} leptogenesis mechanisms. Direct experimental access to particles and LNV interactions at these scales is typically impractical\,\footnote{Long-lived particle searches are accessible in the context of leptogenesis via oscillation of right-handed neutrinos; for example, at the SHiP experiment~\cite{Alekhin:2015byh}.}, due to the small magnitude of the couplings required by the scale of the known, light neutrino masses. Radiative neutrino mass models may accommodate larger, in principle, experimentally accessible, LNV couplings. However, as shown in an effective field theory (EFT) context in Ref.~\cite{Deppisch:2017ecm}, rates for the resulting CP-conserving, LNV processes in the early universe would be so large as to washout the lepton asymmetry before it is converted into the baryon asymmetry (see Ref.~\cite{Harz:2021psp} for a concrete, simplified model realization).

In this work, we show that the extended seesaw (ESS) model~\cite{Kang:2006sn,Mitra:2011qr} provides an exception to these general considerations. Importantly, we find that the ESS model allows one to accommodate the light neutrino masses and low-scale leptogenesis in the presence of experimentally accessible LNV signatures. We focus specifically on searches for $0\nu\beta\beta$ decay,  where the present strongest limit has been set by the KamLAND-Zen experiment on the $^{136}$Xe half-life: $ T_{1/2}^{ 0 \nu} > 2.6 \times 10^{26} $ yrs at 90$\%$ confidence level (C.L.)~\cite{KamLAND-Zen:2022tow}. The upcoming ``ton-scale" $ 0 \nu \beta \beta$-decay searches aim to enhance sensitivity by two orders of magnitude~\cite{Gomez-Cadenas:2019sfa,LEGEND:2017cdu,nEXO:2017nam,Han:2017fol,Armengaud:2019loe,Paton:2019kgy}, potentially detecting signals if the three light neutrinos have Majorana masses and an inverted mass hierarchy spectrum.

We demonstrate that in the ESS model, contributions from additional neutrinos may enhance the $0\nu\beta\beta$-decay rate while maintaining the viability of a low-scale leptogenesis scenario, respectively. The interplay between low-scale leptogenesis and $\onbb$ decay was explored before~\cite{Drewes:2016lqo,Hernandez:2016kel,Asaka:2016zib,Dolan:2018qpy,Chauhan:2021xus,deVries:2024rfh} in both ARS and resonant leptogenesis frameworks. Ref.~\cite{deVries:2024rfh} conducted an EFT analysis of the $0\nu\beta\beta$ decay in the low-scale type-I seesaw model with MeV-GeV sterile neutrinos being assumed. In comparison, we investigate $\onbb$ decay, leptogenesis, and charged lepton flavor violating (cLFV) decay within 
the ESS model, where light sterile neutrinos are naturally required.

As a preview, we highlight key ingredients of the ESS scenario. The model set extends the SM with three additional right-handed neutrinos $(N_R)$, three left-handed sterile neutrinos ($S_L$) that are gauge singlets under the SM gauge group, and a gauge singlet scalar $\phi$. The parameters relevant to the various physical processes are: 
\begin{itemize}
    \item In the context of $ 0 \nu \beta \beta$ decay, the presence of the extra light sterile states will give an additional contribution to the effective mass parameter in comparison to the standard one. The parameters responsible for $ 0 \nu \beta \beta$ decay are the Dirac Yukawa coupling, $Y_D$; the Pontecorvo–Maki–Nakagawa–Sakata  (PMNS) matrix, $U_{\nu}$; light neutrino masses, $\widehat{M}_\nu$; the Casas-Ibarra matrix, $\mathcal{R}$; right-handed neutrino masses, $M_R$; the ESS parameter, $\mu$; and the sterile neutrino masses, $\widehat{M}_S$.  The definition of these parameters is described in detail in Sec.~\ref{sec:model}. These parameters also enter the cLFV decay. Positive signals of $ 0 \nu \beta \beta$ decay are expected in ton-scale $\onbb$-decay experiments for $\mu \lesssim 10\mev$, while the constraints from cLFV searches are always satisfied for $\mu\gtrsim 30$~keV.   
    \item For low-scale leptogenesis, the critically important washout rate depends on $Y_D$, $\mathcal{R}$, $\mu$, $M_R$, the vacuum expectation value (VEV) and mass of the new scalar singlet $v_\phi$ and $M_\phi$, respectively, while the lepton asymmetry depends only on $Y_D$, $M_R$, and $v_\phi$.
    Therefore, there is an additional dependence on  $v_\phi$ and $M_\phi$ in comparison to $0\nu \beta \beta$ decay and cLFV decay. 
    \item We show that the observed baryon asymmetry can be accounted for within the keV-MeV range of the ESS parameter, $\mu$, and $0 \nu \beta \beta$-decay signals are simultaneously detectable in future ton-scale experiments.
\end{itemize}

For a flow chart illustrating the parameter dependence for $ 0\nu\beta\beta$ decay, cLFV decay, and leptogenesis, see Fig.~\ref{fig:parameters}. In Sec.~\ref{sec:conclusion}, we summarize the key ingredients in the model that allow for simultaneously generating the masses of the light, active neutrinos; accommodate non-resonant low-scale leptogenesis; and lead to observable signals future ton-scale $\onbb$-decay searches.

Our discussion of the model and implications in the remainder of the paper is structured as follows: Sec.~\ref{sec:model} delves into the ESS framework, and we discuss the non-unitary and cLFV constraints on the ESS parameters in Sec.~\ref{sec:constraints}. In Sec.~\ref{sec:nldbd}, we proceed to calculate the $0 \nu \beta \beta$ decay parameters, examining various channels and assessing their significance. Sec.~\ref{sec:lepto} is dedicated to exploring leptogenesis, where we compute the CP-asymmetry resulting from the decay of right-handed neutrinos within this model. Finally, we summarize our findings in Sec.~\ref{sec:results}. We conclude in Sec.~\ref{sec:conclusion}.

\section{Framework}
\label{sec:model}
In the extended seesaw model~\cite{Kang:2006sn,Mitra:2011qr}, which is a variation of type-I seesaw model, two sets of gauge-singlet fields $N_{R\alpha}$ ($\alpha=1,2,\ldots,m$) and $S_{L\beta}$ ($\beta=1,2,\ldots,n$) are introduced.
The Lagrangian that describes the neutrino masses is 
\begin{align}
\label{eq:lag}
- \mathcal{L}_{\rm mass} = \dfrac{1}{2}
\begin{pmatrix}
 \bar{\nu}_L, \bar{S}_L, \bar N_R^c
 \end{pmatrix}
\mathcal{M}    
\begin{pmatrix}
\nu_L^c \\
S_L^c \\
N_R
\end{pmatrix}
+ {\rm h.c.}\;,
\end{align}
where we have omitted the flavor indices, ``h.c.'' denotes the Hermitian conjugation, and the neutrino mass matrix $\mathcal{M}$ is defined as
\begin{align}
\mathcal{M} = 
\begin{pmatrix}
0 & 0 & M_D\\
0 & \mu^\prime & M_S\\
M_D^T & M_S^T & M_R
\end{pmatrix}\;.
\label{eq:numassM}
\end{align}

The Dirac mass terms $\bar \nu_L M_D N_R$ and $\bar S_L M_S N_R$ arise from the Yukawa interactions
\begin{align}
-\mathcal{L}_{\rm ESS} &\supset Y_D\, \bar{L}\tilde H N_R + Y_S\, \phi \bar{S}_L N_R + \mathrm{h.c.}\;,
\label{eq:Lag_ESS}
\end{align}
leading to
\begin{align}
M_D = Y_D\, v_H\,,\quad
M_S = Y_S\, v_\phi\,,
\end{align}
where $v_H \equiv \langle H\rangle$ and $v_\phi \equiv \langle \phi\rangle$ denote the VEVs of $H$ and $\phi$, respectively.
The other terms in $\mathcal{M}_\nu$ can also be achieved by introducing new scalar fields and may be incorporated into flipped $SU(5)$ grand unification theories~\cite{Ellis:1992zr, Ellis:1992nq}.

There are two mass scales of lepton number violation, $M_R$, and $\mu^\prime$, in the ESS model. The following assumptions\footnote{To be more precise,
in Eq.~\eqref{eq:hier1} the norm inequality $\left \lVert M_R \right \rVert > \left \lVert M_S \right \rVert$ is required~\cite{Dev:2012sg}, and similar for the other inequalities.} of model parameters are made~\cite{Kang:2006sn,Mitra:2011qr} 
\begin{align}
\label{eq:hier1}
&M_R  > M_S > M_D \gg \mu^\prime\;,\nn\\
&\mu^\prime < M_SM_R^{-1}M_S^T \,,
\end{align}
which ensures the diagonalization of the mass matrix $\mathcal M$. Following Ref.~\cite{Kang:2006sn}, we  
first integrate out the right-handed neutrino fields $N_{R \alpha}$
and obtain the effective Lagrangian as follows
\begin{align}
- \mathcal{L}_{{\rm eff}} = \dfrac{1}{2} 
\begin{pmatrix}
\bar\nu_L& \bar S_L
\end{pmatrix}
M_n
\begin{pmatrix}
\nu_L^c\\
S_L^c
\end{pmatrix} + \textrm{h.c.}\,,
\label{eq:Leff}
\end{align}
where 
\begin{align}
\label{eq:mass_matrix2}
M_n = - \begin{pmatrix}
M_D M_R^{-1} M_D^T  & M_D M_R^{-1} M_S^T\\
M_S M_R^{-1} M_D^T &  M_S M_R^{-1} M_S^T - \mu^\prime
\end{pmatrix}\,.
\end{align} 

We further assume that the numbers of families of right-handed neutrinos $N_R$ and left-handed neutrinos $S_L$ are equal, {\it i.e.,} $m=n$, so the Dirac mass matrix $M_S$ is a square matrix. Moreover, this matrix is assumed to be invertible. 
We diagonalize the mass matrix $M_n$ in two steps~\cite{Mitra:2011qr}. In the first step, $M_n$ is block-diagonalized by the following unitary transformation, 
\begin{align}
\mathcal U_1^\dagger M_n \mathcal U_1^* = 
\begin{pmatrix}
M_\nu^{bd} & 0 \\
0 & M_S^{bd}
\end{pmatrix}\;,\quad
\mathcal U_1 = 
\begin{pmatrix}
\mathbbm{1} & \Theta\\
-\Theta^\dagger & \mathbbm{1}
\end{pmatrix}\;.
\end{align}
It is straightforward to obtain the $3 \times n$ active-sterile neutrino mixing matrix 
\begin{align}
\label{eq:mixing}
\Theta = M_D M_S^{-1} \left[\mathbbm{1} + \mu^\prime \left( M_SM_R^{-1}M_S^T \right)^{-1} \right]
\end{align}
and
\begin{align}
\label{eq:Mnubd}
M_\nu^{bd} & = M_D M_S^{-1} \mu^\prime (M_D M_S^{-1})^T\;,\\
\label{eq:MSbd}
M_S^{bd} & = - M_S M_R^{-1} M_S^T \;.
\end{align}
which agree with Refs.~\cite{Mitra:2011qr,Awasthi-thesis,Jha:2021oxl}. 
For simplicity, we will assume the matrix $\mu^\prime$ to be proportional to the identity, $\mu^\prime = \mu \mathbbm{1}$. It is noted that $\mu^\prime$ is an $n\times n$ matrix, while $\mu$ is a parameter. The active neutrino masses are suppressed by $\mu$ and $M_D M_S^{-1}$, in a manner akin to the inverse seesaw mechanism~\cite{Dolan:2018qpy}, while the sterile neutrino masses can be significantly lower due to the suppression factor $M_S M_R^{-1}$.

In the second step, the block diagonal mass matrix is further diagonalized with another unitary matrix $\mathcal{U}_2$,
\begin{align}
\label{eq:Mdiag2}
\mathcal U_2^\dagger \begin{pmatrix}
M_\nu^{bd} & 0 \\
0 & M_S^{bd}
\end{pmatrix} \mathcal U_2^* = 
\begin{pmatrix}
\widehat M_\nu & 0 \\
0 & \widehat M_S
\end{pmatrix}\;,
\end{align}
where 
\begin{align}
\mathcal U_2 = 
\begin{pmatrix}
U_\nu & 0\\
0 & U_S
\end{pmatrix}\;.
\end{align}
In the above, $U_\nu$ is the PMNS matrix, and $U_S$ is an $n\times n$ matrix that diagonalizes $M_S^{bd}$. The active and sterile neutrino mass matrices are defined as $\widehat M_\nu \equiv {\rm diag}(m_1, m_2, m_3)$ and $\widehat M_S \equiv {\rm diag} (m_4, m_5, \ldots, m_{n+3})$, respectively. Note that, in principle, the mass matrix in Eq.~\eqref{eq:numassM} will receive radiative corrections (see, for instance, Refs.~\cite{Pilaftsis:1991ug,Kersten:2007vk,Roy:2010xq,Lee:2013htl,Lopez-Pavon:2015cga,Moffat:2017feq}) that will not only induce a non-zero $\delta m_\nu^{1-\rm loop}$, but also modify the diagonalization process and the parametrization introduced in the following. While it is possible to ensure the stability of light neutrino masses under radiative corrections for suitable regions of the parameter space~\cite{Drewes:2019byd,Moffat:2017feq,Lopez-Pavon:2015cga}, a comprehensive study beyond leading order goes beyond the scope of our work.

The full neutrino mixing matrix $U$ can be expressed as~\cite{Schechter:1981cv,Korner:1992zk,Grimus:2000vj,Hettmansperger:2011bt}
\begin{align}
U = \mathcal{U}_1 \mathcal{U}_2 = 
\begin{pmatrix}
U_\nu & V \\
T & U_S 
\end{pmatrix}\;,
\label{eq:U}
\end{align}
where
\begin{align}
\label{eq:mixing2}
T = -\Theta^\dagger U_\nu \;,\quad V = \Theta U_S\;.
\end{align}

The mass eigenstates of active neutrinos $\nu_a$ and sterile neutrinos $\nu_s$ are
\begin{align}
\nu_a = \nu_L^\prime + \nu_L^{\prime c} \;,\quad 
\nu_s = S_L^\prime + S_L^{\prime c} 
\;,
\end{align}
with 
\begin{align}
\begin{pmatrix}
\nu_L^\prime \\
S_L^\prime 
\end{pmatrix}
= 
U^\dagger
\begin{pmatrix}
\nu_L\\
S_L
\end{pmatrix} \equiv N_m \;.
\end{align}
Defining $\nu = N_m + N_m^c$, we have $\nu_{ai} = \nu_i$ and $\nu_{si} = \nu_{i+3}$ with $i=1,2,\cdots,n$~\cite{Dekens:2020ttz,deVries:2022nyh}. Without losing generality, we will assume $n=3$ for which the lightest neutrino is massive.

The charged-current interactions can be expressed as
\begin{align}
\label{eq:charged_current}
\mathcal{L}_{CC} = \dfrac{g}{\sqrt{2}} \big[ \bar{\ell}_L \gamma^\mu \nu_L W_{L\mu}^+ + V_{ud} \bar u_L \gamma^\mu d_L W_{L\mu}^+   \big] + {\rm h.c.}\;,
\end{align}
where 
\begin{align}
\nu_L &\equiv (PU) P_L \nu = U_\nu \nu_L^\prime + V S_L^\prime\;
\end{align}
with the projectors~\cite{Dekens:2020ttz} 
\begin{align}
P = (\mathcal{I}_{3\times 3} \quad 0_{3\times 3})\;,\quad
P_s = (0_{3\times 3} \quad \mathcal{I}_{3\times 3} )\;.
\end{align}
In Eq.~\eqref{eq:charged_current}, the element of Cabibbo-Kobayashi-Maskawa matrix $V_{ud}$ describe quark mixing in the first generation,
and $\ell = e,\mu,\tau$ denotes the charged lepton field.

Following Ref.~\cite{Dolan:2018qpy}, we extend the Casas-Ibarra parameterization~\cite{Casas:2001sr} to obtain 
the neutrino Dirac  mass matrix $M_S$ in terms of active neutrino masses and the PMNS matrix, which is
\begin{align}
\label{eq:CI_MS}
M_S &= \mu^{1/2} \mathcal{R}^T \widehat{M}_\nu^{-1/2} U_\nu^\dagger M_D\,.
\end{align}
In the above, $\mathcal R$ is an arbitrary complex orthogonal matrix with $\mathcal R\mathcal R^T=\mathbbm{1}$. The matrix $\mathcal R$ depends on six parameters and can be decomposed as $\mathcal R=O\, H$, where $O$ is an orthogonal matrix, and $H$ is a complex orthogonal, hermitian matrix~\cite{Cirigliano:2006nu, Pascoli:2003rq}. Both $O$ and $H$ can be expressed in terms of three independent parameters,
\begin{align}
H = \exp({\rm i}\,\Phi)\,,\ \Phi &=\begin{pmatrix}
0&\phi_1&\phi_2\\
-\phi_1&0&\phi_3\\
-\phi_2&-\phi_3&0
\end{pmatrix}\,,\\
O = \exp(A)\,,\ A &=\begin{pmatrix}
0&a_1&a_2\\
-a_1&0&a_3\\
-a_2&-a_3&0
\end{pmatrix}\,,
\end{align}
where $\{\phi_i\}$ and $\{a_i\}$ are real parameters. In this parameterization, we obtain that the sterile neutrino masses are proportional to $\mu$, see Eqs.~\eqref{eq:MSbd}, \eqref{eq:Mdiag2}, and \eqref{eq:CI_MS}.

To connect with the leptogenesis,
we further express the sterile neutrino Yukawa couplings as
\begin{align}
Y_S 
= \frac{1}{v_\phi}M_S
= \frac{1}{v_\phi} \mu^{1/2} \mathcal R^T \widehat{M}_\nu^{-1/2} U^\dagger_\nu M_D\;.
\label{eq:CI_YS}
\end{align}

Before closing this section, we would like to comment on the procedure of neutrino mass diagonalization with right-handed neutrino being first integrated out. In this setup, the mixing is described by a $6\times 6$ matrix $U$. The interactions of sterile neutrinos with SM-charged leptons are suppressed by the small $U_{\alpha i} \propto M_D M_S^{-1}$ for $\alpha=e,\mu,\tau$ and $i=4,5,6$. 

Like the sterile neutrinos, the right-handed neutrinos can also contribute to low-energy processes $\mu \to e\gamma$ and $\onbb$ decay. 
To assess their impact, we need to diagonalize the full neutrino mass matrix with a $9\times 9$ mixing matrix (also denoted by $U$). 
Their contributions to the low-energy observables can be incorporated by taking $i=7,8,9$ and identifying $m_7$, $m_8$, and $m_9$ as the masses of three right-handed neutrinos. Nevertheless, as shown in Ref.~\cite{Mitra:2011qr},  $U_{\alpha i} \propto M_D M_R^{-1}$ for $i=7,8,9$, which is smaller than $M_D M_S^{-1}$. This work aims to establish connections among $\onbb$  decay, cLFV decay, and leptogenesis. Therefore, we will omit the contributions of right-handed neutrinos to the low-energy observables.

\section{Non-unitary and cLFV constraints}
\label{sec:constraints}

The ESS parameters are constrained by the neutrino oscillation data and low-energy measurements as the canonical type-I seesaw models~\cite{Abada:2007ux,Dinh:2012bp,Fernandez-Martinez:2016lgt,Blennow:2023mqx}.
From discussions in the previous section, after the diagonalization of the neutrino matrix, one could obtain the active neutrino states $\nu_{ai}$ and sterile neutrinos $\nu_{sj}$ for $i,j=1,2,3$ with the active-sterile mixing described by the mixing matrix $\Theta$. 
Such a non-zero active-sterile mixing implies a deviation 
from the unitarity of the PMNS matrix, which is usually parameterized as $(1-\eta) U_\nu$ with 
$\eta = \Theta \Theta^\dagger$/2~\cite{Blennow:2011vn}. Using the global fits of the flavor and electroweak precision data~\cite{Blennow:2023mqx,Fernandez-Martinez:2016lgt}, $| \eta_{\alpha\beta} | \lesssim 10^{-5}-10^{-3}$ for different flavors 
$\alpha,\beta=e,\mu,\tau$, where the heavy neutrino fields have been integrated out.

However, if the sterile neutrinos are kinematically available in the observables used to derive the constraints, the unitary is restored, and the aforementioned constraint on $|\eta_{\alpha\beta}|$ is not applicable~\cite{Blennow:2016jkn}. For the sterile neutrino mass as low as $\sim 1-10$~eV, the constraint becomes $|\eta_{\alpha\beta}|\lesssim 10^{-2}-10^{-1}$, which has been investigated in Ref.~\cite{Blennow:2016jkn}. From Eqs.~\eqref{eq:mixing},~\eqref{eq:CI_MS}, $\Theta \sim (\widehat M_\nu/\mu )^{1/2} $, we thus obtain rather weak constraint $\mu \gtrsim 1$~eV with the other ESS parameters being specified in Tab.~\ref{tab:param_scan}.

Besides the impact on neutrino oscillation, the charged current interactions in Eq.~\eqref{eq:charged_current} would give rise to the cLFV decays.
The branching ratio for the radiative decay $\mu \rightarrow e \gamma $ is given by~\cite{Ilakovac:1994kj}
\begin{align}
{\text{BR}} (\mu \rightarrow e \gamma) &= \frac{G_F^3 \sin^2 \theta_W}{64\sqrt{2}\, \pi^5} \dfrac{m_W^2\,m_\mu^5}{\Gamma_\mu} 
|G_{\mu e}|^2\;.
\label{eq:LFV}
\end{align}
Here, 
$G_F$ is the Fermi constant,
$\theta_W$ is the weak mixing angle, and $m_{\mu}$, 
$m_W$ are the masses of the muon and $W$ boson, respectively. 
Using the expression of the total decay width of muon $\Gamma_{\mu}$, we have~\cite{Dinh:2012bp,Cheng:1980tp}
\begin{align}
\label{eq:g-factor}
 {\text{BR}} (\mu \rightarrow e \gamma)   & \simeq \dfrac{3}{2\pi} \alpha|G_{\mu e}|^2\;,
\end{align}
where $\alpha$ is the fine structure constant, and
\begin{align}
G_{\mu e} = \sum_{i=1}^6 U_{\mu i}^* U_{ei}  G_\gamma \left(\frac{m_{i}^2}{m_W^2} \right)
\end{align}
with the loop function being defined as
\begin{align}
    & G_\gamma(x) = - \frac{2 x^3 + 5 x^2 -x}{4(1-x)^3} - \frac{3 x^3 \text{ln} x}{2(1-x)^4}\;.
    \label{eq:loopfactor}
\end{align}
For $x\ll 1$, $G_\gamma(x) \to x/4$, while for $x\gg 1$, $G_\gamma(x) \to 1/2$. In the latter case, the factor $|U_{\mu i}^* U_{ei}|$ is  suppressed by the sterile neutrino mass~\cite{Blennow:2023mqx,Fernandez-Martinez:2016lgt}.

The most recent experimental bound given by the MEG II experiment is ${\rm BR}(\mu \to e \gamma) < 3.1 \times 10^{-13}$ at 90\% C.L.~\cite{MEGII:2023ltw},
while the future sensitivity is expected to be $\text{BR} (\mu \to e \gamma) < 6 \times 10^{-14}$~\cite{MEGII:2018kmf}.  Details of the constraints on the model parameters will be given in Sec.~\ref{sec:results}.

We also comment on the constraints from other cLFV observables.
In the ESS model, the rate of $\mu-e$ conversion and the branching ratio of $\mu \to eee$ are strongly correlated with the branching ratio of $\mu \to e\gamma$. This is because the photonic contribution dominates over the other contributions in $\mu-e$ conversion and $\mu \to eee$ processes, similar to the case in the inverse seesaw model~\cite{Dolan:2018qpy,Deppisch:2005zm}. The ratios are given by~\cite{Kuno:1999jp}
\begin{align}
\label{eq:correlation}
\dfrac{{\rm CR}(\mu N \to e N)}{{\text{BR}} (\mu \rightarrow e \gamma)} &= \dfrac{3\cdot 10^{12} G_F^2 m_\mu^4}{96\pi^3 \alpha} B(A,Z)\;,\nn\\
\dfrac{{\rm BR}(\mu \to eee)}{{\text{BR}} (\mu \rightarrow e \gamma)} &= \dfrac{\alpha}{3\pi} \left( \ln\dfrac{m_\mu^2}{m_e^2}-\dfrac{11}{4} \right)\;,
\end{align}
where $B(A,Z)$ represents the rate dependence on the target nucleus with the mass number $A$ and the atomic number $Z$, and $m_e$ denotes the mass of the electron, which is about $1\sim 2$ for different target nucleus~\cite{Kuno:1999jp,Kitano:2002mt}.
Numerically, one obtains ${\rm CR}(\mu N \to e N)/{\text{BR}} (\mu \rightarrow e \gamma)\simeq B(A,Z)/428$ and ${\rm BR}(\mu \to eee)/{\text{BR}} (\mu \rightarrow e \gamma) \simeq 6\times 10^{-3}$. 
In comparison with the current experimental constraints, ${\rm BR}(\mu \to eee) < 1.0 \times 10^{-12}$ (SINDRUM~\cite{SINDRUM:1987nra}), ${\rm CR}(\mu {\rm Au} \to e {\rm Au}) < 7\times 10^{-13}$ (SINDRUM II~\cite{SINDRUMII:2006dvw}), ${\rm CR}(\mu {\rm Ti} \to e {\rm Ti}) < 6.1\times 10^{-13}$ (SINDRUM II~\cite{Wintz:1998rp}), we obtain that the cLFV decay $\mu\to e\gamma$ gives the most stringent constraint than the other cLFV processes. The future sensitivities of ${\rm BR}(\mu \to eee) $ and ${\rm CR}(\mu {\rm Ti} \to e {\rm Ti})$ could potentially reach $\sim 10^{-16}$ (Mu3e~\cite{Blondel:2013ia}) and $\sim 10^{-17}$ (Mu2e~\cite{Mu2e:2008sio}, COMET~\cite{COMET:2018wbw}), respectively. Due to the strong correlation, we can compare their sensitivities by evaluating $G_{\mu e}$ defined in Eq.~\eqref{eq:g-factor}. Future limits on $|G_{\mu e}|$ from the experimental searches for $\mu\to eee$ and $\mu -e$ conversion with the ultimate sensitivities as quoted are expected to be $\sim 1.9$ and 6 times stronger than that for $\mu \to e\gamma$, respectively. Our results are consistent with the findings in Ref.~\cite{Haxton:2022piv} with delicate studies in the EFTs.

\section{\tf{$0 \nu \beta \beta$}{0vbb} decay with sterile neutrinos}
\label{sec:nldbd}

This section focuses on the calculation of $0 \nu \beta \beta$-decay half-lives across a wide spectrum of neutrino masses.
Within the context of the ESS model, we 
consider the contributions to $\onbb$-decay half-life from the exchange of sterile neutrinos $\nu_{si}$ with $i=1,2,3$, the masses of which 
could be of the order of the right-handed scale or
significantly lower.
Given this hierarchy of energy scales, it is advantageous to employ the EFT. Refs.~\cite{Prezeau:2003xn,Cirigliano:2017djv,Cirigliano:2018yza,Dekens:2020ttz} have devised a comprehensive EFT framework that links high-scale sources of lepton number violation to $0 \nu \beta \beta$-decay rates.

In the presence of light sterile neutrinos, we adopt the so-called
neutrino-extended SM EFT framework~\cite{Lehman:2014jma, Liao:2016qyd}, which encompasses all effective operators that are invariant under the SM gauge group, containing both SM fields as well as sterile neutrino fields. Subsequently, the higher-dimensional operators are evolved to the weak scale, at which point heavy SM fields are integrated out. At even lower energy levels ($\Lambda_\chi\sim 1$ GeV), the resultant EFT operators are coupled with hadronic LNV operators through the application of chiral EFT, which represents the low-energy EFT of QCD.

We integrate out the $W$ boson at the weak scale and match with an effective Lagrangian that preserves $SU(3)_C \times U(1)_{\rm em}$ symmetry as follows~\cite{Dekens:2020ttz}:
\begin{align}
\mathcal L^{(6)}_{\nu \text{EFT}} & \supset \frac{2 G_F}{\sqrt{2}} \bar u_L \gamma_\mu d_L \bar e_L \gamma^\mu C^{(6)}_{{\rm VLL}} \nu  \;.
\end{align}
Here,
the Wilson coefficient 
\begin{align}
\label{eq:WC}
&\left(C_{\mathrm{VLL}}^{(6)}\right)_{e i}=-2 V_{u d}(P U)_{e i}
\nn\\
&=
\begin{cases}
-2 V_{u d}(U_\nu)_{e i}, & \text{for $i=1,2,3$}\,,\\
-2 V_{u d}(\Theta U_S)_{e (i-3)}, & \text{for $i=4,5,6$}\,,
\end{cases}
\end{align}
where $\Theta$ is defined in Eq.~\eqref{eq:mixing}.

The inverse half-life is expressed as~\cite{Dekens:2020ttz,Cirigliano:2018yza}
\begin{equation}
\begin{aligned}
\label{eq:half-life}
\left(T_{1 / 2}^{0 \nu}\right)^{-1}= g_A^4 G_{01}\left| \sum_{i=1}^6 \mathcal{A}_L\left(m_i\right) \right|^2\;,
\end{aligned}
\end{equation}
where $G_{01}$ is the phase space factor, $g_A= 1.27$, and the sub-amplitude 
\begin{align}
\label{eq:subamplitude}
\mathcal{A}_L\left(m_i\right)= & -\frac{m_i}{4 m_e} \left(C_{\mathrm{VLL}}^{(6)}\right)_{e i}^2\bigg[ \mathcal{M}_V + \mathcal{M}_A \nn\\ 
&\qquad+ \dfrac{2m_\pi^2}{g_A^2} g_\nu^{NN}\left(m_i\right)  M_{F, s d}\bigg] \;. 
\end{align}
Here, 
$m_\pi$ and $m_e$ are the masses of pion and electron, respectively. 
$\mathcal{M}_V$ and $\mathcal{M}_A$ are defined as combinations of nuclear matrix elements (NMEs):
\begin{align}
\mathcal{M}_V & =-\frac{1}{g_A^2} M_F\left(m_i\right)+M_{G T}^{M M}\left(m_i\right)+M_T^{M M}\left(m_i\right)\,, \\
\mathcal{M}_A & = M_{G T}^{A A}\left(m_i\right)+M_{G T}^{A P}\left(m_i\right)+M_{G T}^{P P}\left(m_i\right)\nn\\
&\quad +M_T^{A P}\left(m_i\right)+M_T^{P P}\left(m_i\right)\,.
\end{align}

The subscripts correspond to the Fermi $(F)$, Gamow-Teller $(GT)$, and Tensor $(T)$ contributions to the nuclear matrix elements, whereas the superscripts refer to axial $(A)$, pseudoscalar $(P)$, and magnetic $(M)$. The mass-dependent NMEs $M_X^Y (m_i)$ with $X=F,GT,T$ and $Y=AA,AP,PP,MM$ can be expressed using interpolation formulae\,\footnote{
An improved treatment, including the contribution of ultrasoft neutrinos, was recently proposed in Ref.~\cite{Dekens:2024hlz}.
}~\cite{Faessler:2014kka,Dekens:2020ttz}. For instance, the Fermi NME is
$M_F(m_i ) = m_\pi^2 \left[ m_i^2 + m_\pi^2 M_{F,sd}/M_F \right]^{-1} M_{F,sd}$.
The reader can refer to Ref.~\cite{Dekens:2020ttz} for the remaining expressions. Notice that the NMEs $M_X^Y (m_i)$ have the limits $M_X^Y$ and $M_{X,sd}^Y$ if $m_i \ll m_\pi$ and $m_i \gg m_\pi$. The values of NMEs $M_X^Y$ and $M_{X,sd}^Y$ for isotopes calculated in different nuclear methods can be found in Ref.~\cite{Cirigliano:2018yza}.
Similarly, the mass-dependent low-energy constant $g_\nu^{NN}\left(m_i\right)$ is also derived using an interpolation formula~\cite{Dekens:2020ttz}.

If $m_i\ll m_\pi$, as seen in Eq.~\eqref{eq:subamplitude}, $\mathcal{A}_L(m_i) \propto U_{ei}^2 \,m_i$.
If $m_i\gg m_\pi$, however, $\mathcal{M}_V + \mathcal{M}_A \propto 1/m_i^2$, so that $\mathcal{A}_L(m_i) \propto U_{ei}^2/m_i$. 
The amplitude in the large $m_i$ limit agrees with that obtained in the description of dimension-9 LNV operators~\cite{Prezeau:2003xn}. In both cases, the sub-amplitudes of the sterile neutrinos are suppressed by $U_{ei}^2 = \left(\Theta U_S\right)_{e(i-3)}^2$ $(i=4,5,6)$ with the active-sterile mixing 
$\Theta \sim (\widehat M_\nu/\mu )^{1/2}$. Thus we obtain that the sub-amplitude of the sterile neutrino with mass $m_i$ is a factor of $m_{i}/\mu$  compared to the sub-amplitudes of active neutrinos in case of $m_{i} \ll m_\pi$, while it is a factor of $m_{\pi}^2/(\mu m_{i})$ in case of $m_{i} \gg m_\pi$.

For our numerical analysis, we will consider $\onbb$-decay searches in $^{136} \mathrm{Xe}$, for which the phase space factor $G_{01}=1.5\times 10^{-14}$ yrs$^{-1}$~\cite{Cirigliano:2018yza}.
The most stringent constraint  $T^{0\nu}_{1/2} \left(^{136}{\rm Xe}\right) > 2.6 \times 10^{26} {\rm\ yrs}$ at 90\% C.L. is obtained by KamLAND-Zen experiment~\cite{KamLAND-Zen:2022tow}, while 
future prospect $T^{0\nu}_{1/2} \left(^{136}{\rm Xe}\right) > 10^{28} {\rm\ yrs}$ is expected in 
the ton-scale experiments~\cite{Han:2017fol,Gomez-Cadenas:2019sfa,LEGEND:2017cdu,nEXO:2017nam}. 
The sensitivities to the model parameter $\mu$ will be presented in Sec.~\ref{sec:results}.

\section{Leptogenesis with right-handed and sterile neutrinos}
\label{sec:lepto}

Beyond its elegant explanation for the small scale of neutrino masses, the seesaw mechanism also undergirds the leptogenesis framework~\cite{Fukugita:1986hr}. However, conventional thermal leptogenesis requires the mass of the lightest right-handed neutrino to be larger than $\sim 10^9$ GeV with a hierarchical mass spectru~\cite{Davidson:2002qv}, rendering it unfeasible to explore thermal leptogenesis directly in collider experiments anytime soon. Additionally, thermal leptogenesis at such a high mass scale is hindered by the gravitino problem within the framework of the supersymmetric SM~\cite{Khlopov:1984pf,Campbell:1992hd,Ross:1995dq,Sarkar:1995du,Giudice:1999am}. In contrast, low-scale ({\it e.g.}, TeV and below) leptogenesis scenarios can be experimentally accessible while avoiding the gravitino problem. One example is resonant leptogenesis, which is characterized by the additional Majorana neutrinos having a relatively small mass splitting~\cite{Hambye:2001eu,Pilaftsis:2003gt,DeSimone:2007edo,Hambye:2004jf,Pilaftsis:2005rv,Cirigliano:2006nu,Xing:2006ms,Branco:2006hz,Iso:2010mv,Iso:2013lba,BhupalDev:2014pfm,Aoki:2015owa,Dev:2017wwc,Asaka:2018hyk,Brivio:2019hrj,Brdar:2019iem,Mohanty:2019drv}. Similarly, in ARS leptogenesis, the right-handed neutrinos are at the electroweak scale, and the CP-violating effects are induced by the coherent superposition of different right-handed mass eigenstates~\cite{Drewes:2017zyw, Akhmedov:1998qx}. Extensions of these mechanisms have also been proposed to achieve successful low-scale leptogenesis ---see, for example, Refs.~\cite{Kang:2006sn, Hugle:2018qbw, Fischer:2021nha, Singh:2023eye, Garbrecht:2024bbo}.

\begin{figure}[htb!]
\centering
\includegraphics[scale=0.40]{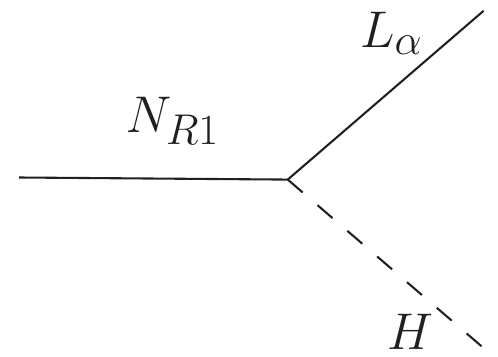} \qquad
\includegraphics[scale=0.33]{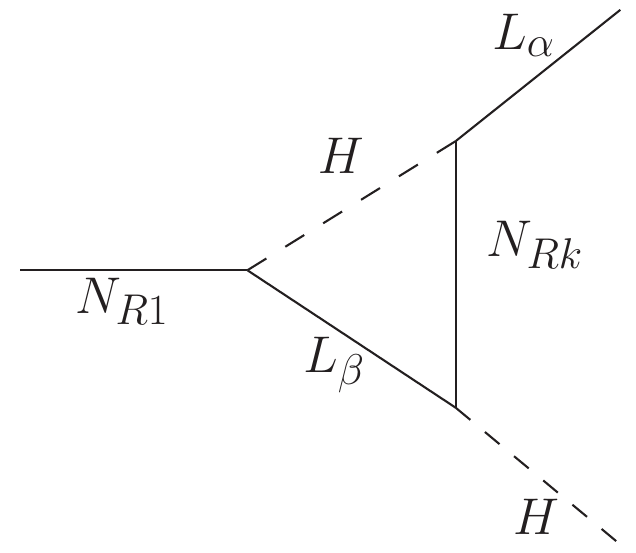} \qquad
\includegraphics[scale=0.35]{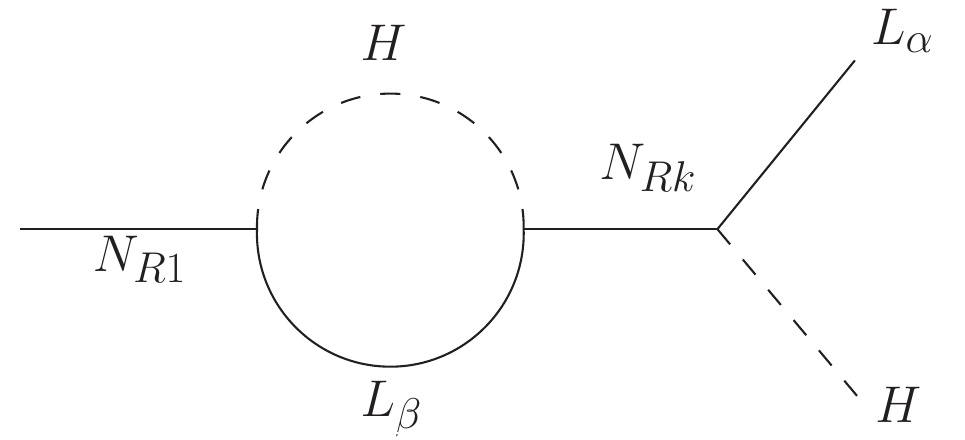} \qquad
\includegraphics[scale=0.35]{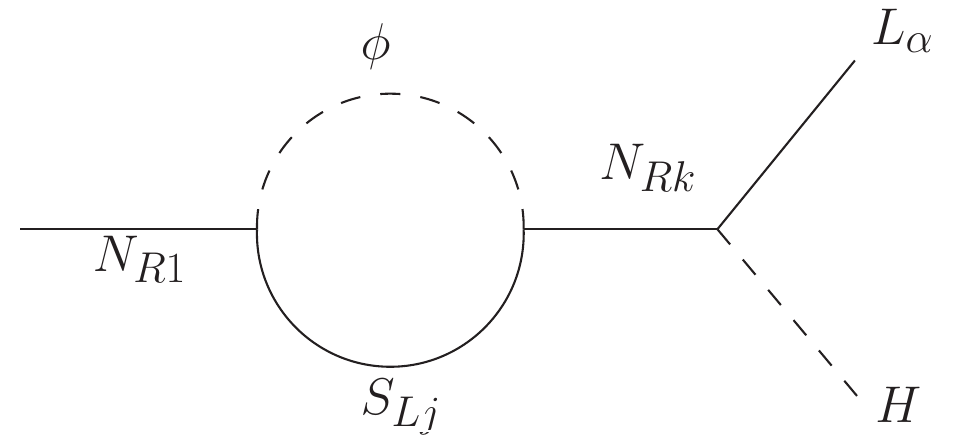}
\caption{
Diagrams contributing to the lepton asymmetry in the ESS model: tree-level decay of $N_{R1}$, vertex correction, and self-energy diagrams. The first two rows correspond to the usual contributions in thermal leptogenesis, whereas the third row is an entirely new contribution due to sterile neutrinos.
}
\label{fig:LGdecay}
\end{figure}

Here, we apply the low-scale leptogenesis framework proposed in Ref~\cite{Kang:2006sn} to the ESS model. In contrast with standard thermal leptogenesis, the neutrino mass relation in Eq.~\eqref{eq:Mnubd} and the additional contributions to the CP-asymmetry due to the presence of the sterile neutrinos (last row in Fig.~\ref{fig:LGdecay}) allow the ESS model to circumvent the Davidson-Ibarra bound.
 
In principle, one needs to solve three coupled Boltzmann equations. Assuming that $N_{R1}$ and $N_{R2}$ are nearly mass-degenerate right-handed neutrinos and their masses are very much smaller than $N_{R3}$, as well as assuming the two of them have approximately the same interactions, so their comoving density satisfies $Y_{N_{R1}} \approx Y_{N_{R2}}$, we simultaneously solve the Boltzmann equation for the lightest right-handed neutrino comoving density, $Y_{N_{R1}}$ and the $B-L$ asymmetry, $Y_{B-L}$. For simplicity, we assume all the SM particles and the new scalar and left-handed sterile neutrinos are in equilibrium. The comoving density is defined as the ratio of the actual number density to the entropy density of the universe.

The CP-asymmetry associated with the decays of the lightest right-handed neutrino $N_{R1}$ to leptons $L_\alpha$ $(\alpha=e,\mu,\tau)$ and SM Higgs boson $H$ is
\begin{equation}
    \epsilon_1 = \frac{\sum_\alpha \left[ \Gamma(N_{R1} \rightarrow L_\alpha H) - \Gamma(N_{R1} \rightarrow \bar{L}_\alpha H^\dag) \right]}{\sum_\alpha \left[ \Gamma(N_{R1} \rightarrow L_\alpha H) + \Gamma(N_{R1} \rightarrow \bar{L}_\alpha H^\dag) \right]}\,.
\end{equation}

It is noted that the asymmetry for $N_{R_2}$ is identical with the substitution $1\to 2$ in the above equation.

The diagrams contributing to $\epsilon_1$ are shown in Fig.~\ref{fig:LGdecay} and the analytical expression is given in the Appendix~\ref{app:leptogenesis} in Eq.~\eqref{eq:eps_full}. Following Ref.~\cite{Kang:2006sn}, we employ the perturbative expression for the CP asymmetry in the ESS low-scale scenario,
\begin{align}
\epsilon_1 \simeq \frac{1}{8\pi} \sum_{k \neq 1} & \Bigg\{\frac{\text{Im}\left[(Y_D^\dagger Y_D)^2_{k1} + (Y_D^\dagger Y_D + Y_S^\dagger Y_S)_{1k}\right]}{(Y_D^\dagger Y_D + Y_S^\dagger Y_S)_{11}}\nonumber\\
& \left(\frac{1}{1-x_k}\right)\Bigg\}\,
\label{eq:eps_res}
\end{align}
with $x_k = M_{Rk}^2/M_{R1}^2$ for $k \neq 1$. For simplicity, henceforth, we will consider that the effects of $N_3$ are negligibly small, i.e., $M_{R1} \simeq M_{R2} \ll M_{R3}$.

\begin{figure}[htb!]
\centering
\includegraphics[scale=0.35]{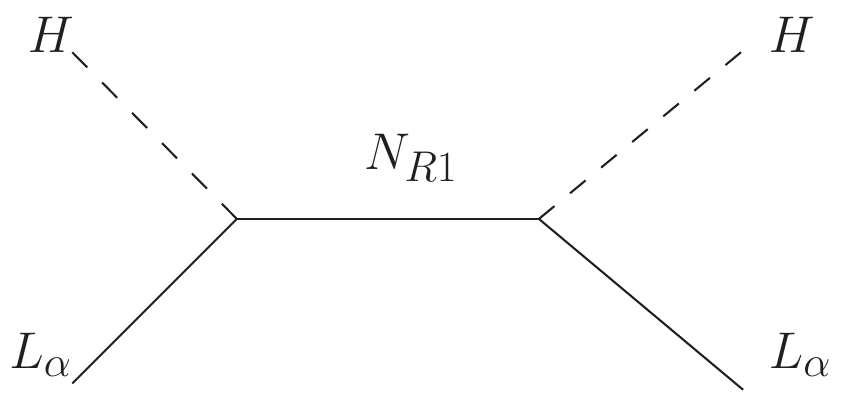} \qquad
\includegraphics[scale=0.35]{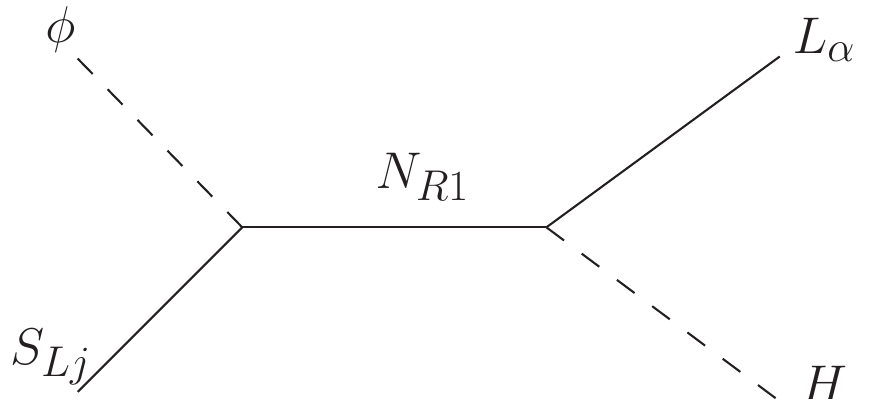}
\includegraphics[scale=0.28]{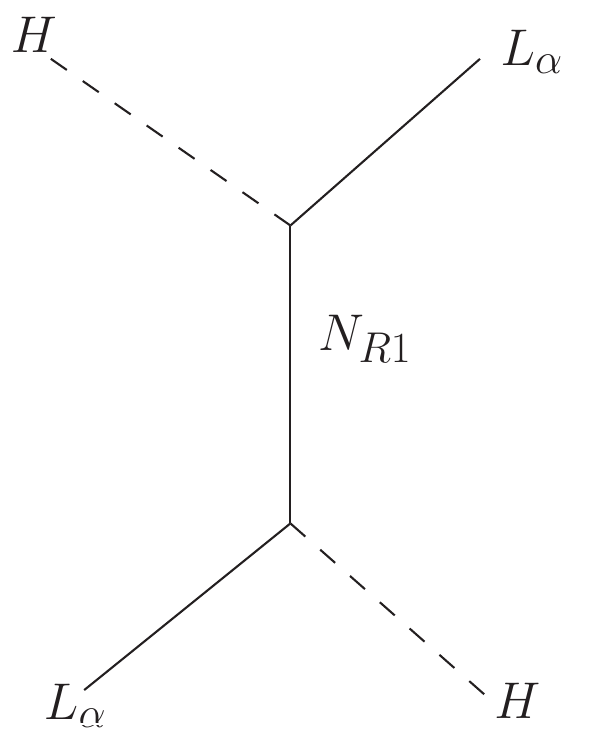}
\caption{$\Delta L = 2$ $s$-channel and $t$-channel scattering process via $N_{R1}$; a new contribution, in addition to the standard leptogenesis scenario, involves the scalar $\phi$ in the initial state.
}
\label{fig:LGscattering1}
\end{figure}

\begin{figure}[htb!]
\centering
\includegraphics[scale=0.35]{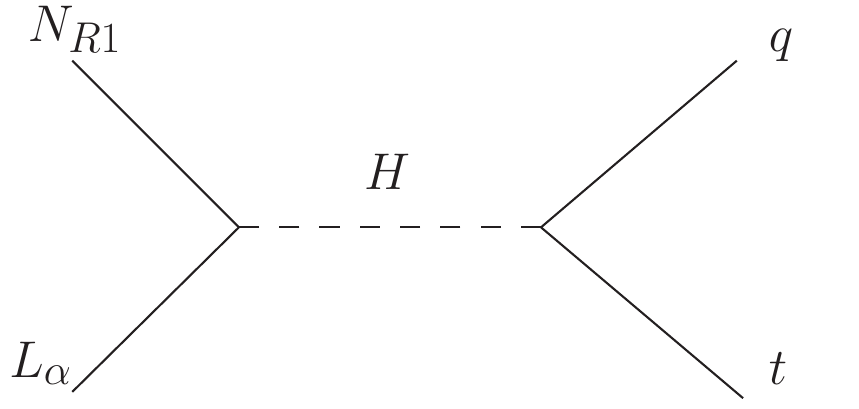} \qquad
\includegraphics[scale=0.30]{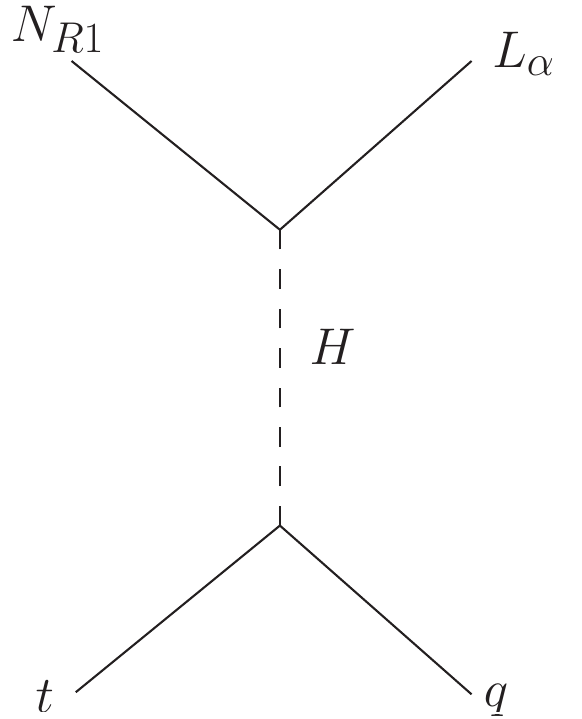} \qquad
\includegraphics[scale=0.30]{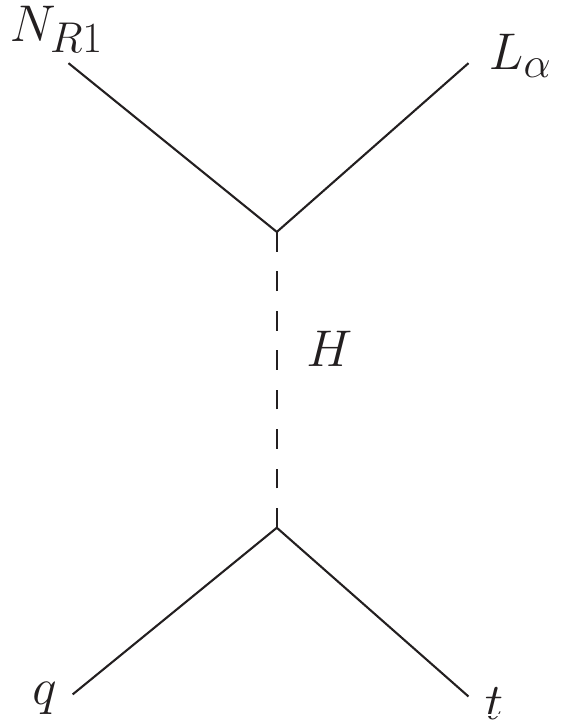}
\caption{$\Delta L = 1$ $s$-channel and $t$-channel scattering process via Higgs exchange. 
}
\label{fig:LGscattering2}
\end{figure}

In general, the Boltzmann equations for $Y_{N_{R1}}$ and $Y_{B-L}$ are formulated to include the decay/inverse decay rates and the scattering cross sections~\cite{Plumacher:1996kc}. The complete set of equations is shown in Appendix~\ref{app:leptogenesis}. The ESS model considers LNV scattering diagrams involving the lightest right-handed neutrino shown in Figs.~\ref{fig:LGscattering1}~and~\ref{fig:LGscattering2}. Notice that there are extra contributions to the $\Delta L = 2$ processes in addition to the standard leptogenesis scenario. These additional contributions arise from diagrams involving the singlet scalar $\phi$ in the initial state. These new interactions significantly modify the washout rate, which differs from the standard case. The detailed analytical expressions for these washout contributions within this model can be found in Appendix~\ref{app:washout}. At a temperature $T \gtrsim M_{R1}$, we solve these two equations assuming zero comoving densities as initial conditions. The particular choice of the initial conditions, either zero or thermal equilibrium densities, does not significantly affect the final value of the asymmetry.

The lepton asymmetry is converted into baryon asymmetry, resulting in the final baryon number at the freeze-out temperature of the sphaleron process, \textit{i.e.}, $T_{\text{sph}} = 131.7$ GeV~\cite{DOnofrio:2014rug} as pointed out in Ref.~\cite{Burnier:2005hp},
\begin{equation}
    Y_{B} = \bigg( \frac{8 N_{f} + 4 N_{H}}{22 N_{f} + 13 N_{H} } \bigg) Y_{B-L} (z_{\text{sph}})\,,
\end{equation}
where $N_f$ and $N_H$ are the number of generations of fermion families and the number of Higgs doublets, respectively. In our scenario, $N_f = 3$ and $N_H =1$.

In our model, the Dirac Yukawa couplings $Y_D$ and $Y_S$ in Eq.~\eqref{eq:new_cs} are related through the Casas-Ibarra parametrization. Within the ESS model, the washout rate is influenced by several parameters: $Y_D$, $R$, $\mu$, $M_R$, $v_\phi$, and $M_\phi$. Meanwhile, the generation of lepton asymmetry primarily depends on $Y_D$ and $M_R$. Notably, the dependence on $v_\phi$ and $M_\phi$ introduces additional factors that distinguish this model from predictions for $0\nu\beta\beta$ and cLFV observables. In general, large values of $Y_D$ and $Y_S$ would typically result in excessively large washout rates, inhibiting the generation of the correct BAU. 

However, in the standard scenario as shown in Fig.~\ref{fig:LGscattering2}, when the temperature drops below $M_{R1}$ (i.e., when $z > 1$), the washout effects become Boltzmann suppressed as $N_{R1}$ 
transits to a non-relativistic state. As the universe continues to cool down and the temperature falls further below $M_{R1}$, the kinematic conditions increasingly disfavor the washout processes involving $N_{R1}$ since $N_{R1}$ can no longer be produced on the shell. Consequently, as $z$ increases, both the Boltzmann suppression of the right-handed neutrino density and the decoupling of $N_{R1}$ from the plasma work together to reduce the effectiveness of the washout. In our model, despite the large value of $Y_D$, Boltzmann suppression remains significant when $T < M_{R1} = 10$ TeV, ensuring the washout remains low in the standard case. 
The standard contribution depicted in Fig.~\ref{fig:LGscattering1} includes an off-shell propagator effect, which is suppressed by the mass of $N_{R1}$. Moreover, the new diagrams involving $ \phi $ and $S_L$ introduce an additional contribution of $({v_H}/{v_\phi}) \mu^{1/2} R^T \widehat{M}_\nu^{-1/2} U_\nu^\dagger$, alongside the suppression by the heavy $N_{R1}$, relative to the conventional scenario.
To maintain a low washout effect, $v_\phi$ must be sufficiently large to effectively suppress this contribution. Moreover, the model allows for a large CP asymmetry due to the substantial $Y_D$, which opens up the possibility for low-scale leptogenesis at TeV scales.


\section{Constraints from BBN and Direct Searches}
\label{sec:bbn_&_direct}

Both cosmological Big Bang Nucleosynthesis (BBN) implications and direct searches for sterile neutrinos impact the allowed parameter space by constraining the mixing angles as a function of the sterile neutrino mass; see, for instance, Refs.~\cite{Bolton:2019pcu,Fernandez-Martinez:2023phj}.

In principle, light sterile neutrinos are strongly constrained by BBN~\cite{Dolgov:2003sg}, as additional relativistic degrees of freedom can modify the early-Universe expansion rate and impact light-element abundances. Nevertheless, these bounds can be relaxed in various non-standard cosmological scenarios. For example, late-time entropy production~\cite{Fuller:2011qy} can dilute the sterile neutrino abundance after decoupling, while the existence of additional light states, such as axion-like particles~\cite{Deppisch:2024izn}, can open alternative decay channels that reduce their cosmological effects. Such mechanisms therefore allow regions of the light sterile neutrino parameter space to remain compatible with cosmological observations. Since our analysis does not make any specific assumption about these non-standard scenarios, we have not imposed these constraints, though we include this discussion for completeness.

On the other hand, the active-sterile mixing element
$U_{\alpha i} = (\Theta U_S)_{\alpha (i-3)}$ $(i=4,5,6)$
is directly constrained by meson-decay and beam-dump experiments. 
These direct-search constraints are imposed on the production and/or decay of the sterile neutrinos, and are thus not weakened by possible cancellations in the neutrino mass matrix. 
For the purpose of our estimate, we take the representative bound
$|U_{\alpha i}|^2 \lesssim 10^{-4}$
from meson-decay searches in PIENU~\cite{PIENU:2017wbj} and NA62~\cite{NA62:2020mcv}. 
Using $\Theta \sim (\widehat M_\nu/\mu)^{1/2}$ derived in Sec.~\ref{sec:constraints}, this bound implies
$\mu \gtrsim 0.5~{\rm keV}$
for $\widehat M_\nu \sim 0.05~{\rm eV}$. This indicative estimate therefore serves as a useful guide for assessing the phenomenological viability of parameter regions with small $\mu$.

\section{Result and Discussion}
\label{sec:results}

\begin{figure}[htb!]
\centering
\includegraphics[width=0.45\textwidth]{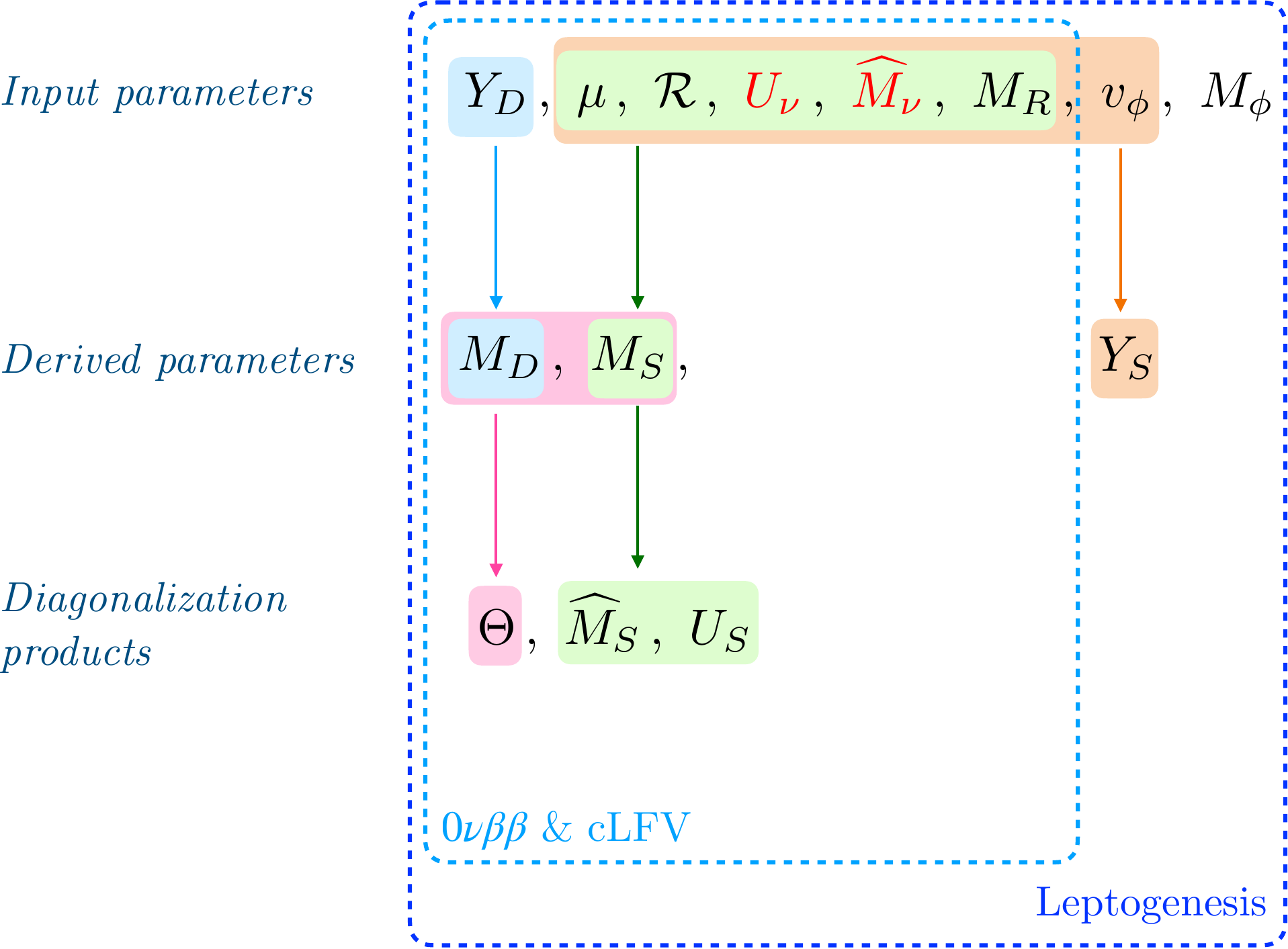} 
\caption{Flow chat for the parameters in the ESS model. The first row corresponds to the input parameters, as shown in Tab. \ref{tab:param_scan}. Among these parameters, the parameters in red ($U_{\nu}$ and $\widehat{M}_\nu$) are either known or constrained. The second and third rows are derived quantities, either by direct definitions or after a diagonalization process. The $\onbb$ decay and cLFV decay rates
dependence, as well as for leptogenesis, is also shown. 
}
\label{fig:parameters}
\end{figure}

We now discuss the results for cLFV decay, $\onbb$ decay, and leptogenesis. To facilitate a clear description of the parameter-dependence for these processes, we refer to the \lq parameter flow chart\rq\, in Fig.~\ref{fig:parameters}. The $\onbb$ decay and cLFV decay rates
depend on the parameters $Y_D$, $M_R$, $\mathcal{R}$ and $\mu$ while leptogenesis carries an additional dependence on 
$v_\phi$ and $M_\phi$. For the implications on $0 \nu \beta \beta$ decay, in light of the discussion in Sec.~\ref{sec:lepto} for low-scale leptogenesis, we consider the two right-handed neutrinos $N_{R1,2}$ with masses at the TeV scale and the third right-handed neutrino to be much heavier. 

\subsection{cLFV decay and \tf{$\onbb$}{0vbb} decay}

\begin{table}[hbt!]
\centering
\begin{tabular}{lrr}
\toprule
\textbf{Parameter} & \textbf{Range} & \textbf{\ Distribution}\\
\midrule
\textit{Casas-Ibarra matrix $\mathcal R$}\\
$a_{i}$ & $[-10,10]$ & flat \\
$\phi_{i}$ & $[-10, 10]$ & flat\\ \\
\textit{Active neutrinos}\\
$\lambda_1$, $\lambda_2$ [rad] & $[0, 2\pi]$ & flat\\ 
$m_{\nu {\rm min}}$ [eV] & $[10^{-5},0.05]$ & log \\ \\
\textit{Sterile neutrinos}\\
$\mu$ [GeV] & $[10^{-8},1]$ & log\\ \\
\textit{Right-handed neutrinos}\\
$M_{R1}$ [TeV] & $[1,50]$ & flat \\
$\Delta M_{21}$ [TeV] & $[10^{-3},10]$ & log \\
$M_{R3}$ [TeV] & $[10^3,10^{12}]$ & log \\
$|Y_D^{ij}|$ & $[10^{-6},1]$ & log\\
$\theta_D^{ij}$ & $[-\pi,\pi]$ & flat\\ \\
\textit{Scalar singlet}\\
$v_{\phi}$ [GeV] & $[500,10^4]$ & flat \\
$M_{\phi}$ [TeV] & $2$ & ---\\
\bottomrule
\end{tabular}
\caption{Parameter ranges adopted for the model scans. $\lambda_{1,2}$ are the Majorana phases, while the rest of the active neutrino parameters are taken from Ref.~\cite{Esteban:2020cvm}, where $m_{\nu {\rm min}}=m_1$ in the NH and $m_3$ in the IH. The mass splitting 
is defined as $M_{R2}\equiv M_{R1}+\Delta M_{21}$. 
The entries of the Yukawa matrix $Y_D$ are parametrized by $Y_D^{ij}=|Y_D^{ij}|\exp ({\rm i}\,\theta_D^{ij})$.
}
\label{tab:param_scan}
\end{table}

In order to investigate the sensitivities of cLFV decay and $\onbb$ decay, we perform a numerical scan for $10\, \mathrm{eV}\, \leq\mu\leq 1\, \mathrm{GeV}$, assuming a flat distribution. We note that the region $\mu\lesssim 0.5\,\mathrm{keV}$ is retained only for completeness in the exploratory scan and is subject to the corresponding active-sterile mixing constraints.
Neutrino oscillation mixing parameters are taken from Ref.~\cite{Esteban:2020cvm}. Both cases of active neutrino mass hierarchy are investigated in this work.
The full set of parameters is varied within the ranges shown in Tab.~\ref{tab:param_scan}, where the mass of singlet scalar $\phi$ is set to be $2\tev$. The assumed conditions in Eq.~\eqref{eq:hier1} are imposed as theoretical cuts during the scanning process, and perturbativity constraints~\cite{Allwicher:2021rtd} for the Yukawa couplings $Y_S^{ij}$ are imposed. We find that after mass diagonalization, the sterile neutrinos may have masses ranging from $10^{-4}$~eV to $10^3$~TeV.

\begin{figure}[htb!]
\centering
\includegraphics[width=0.8\linewidth]{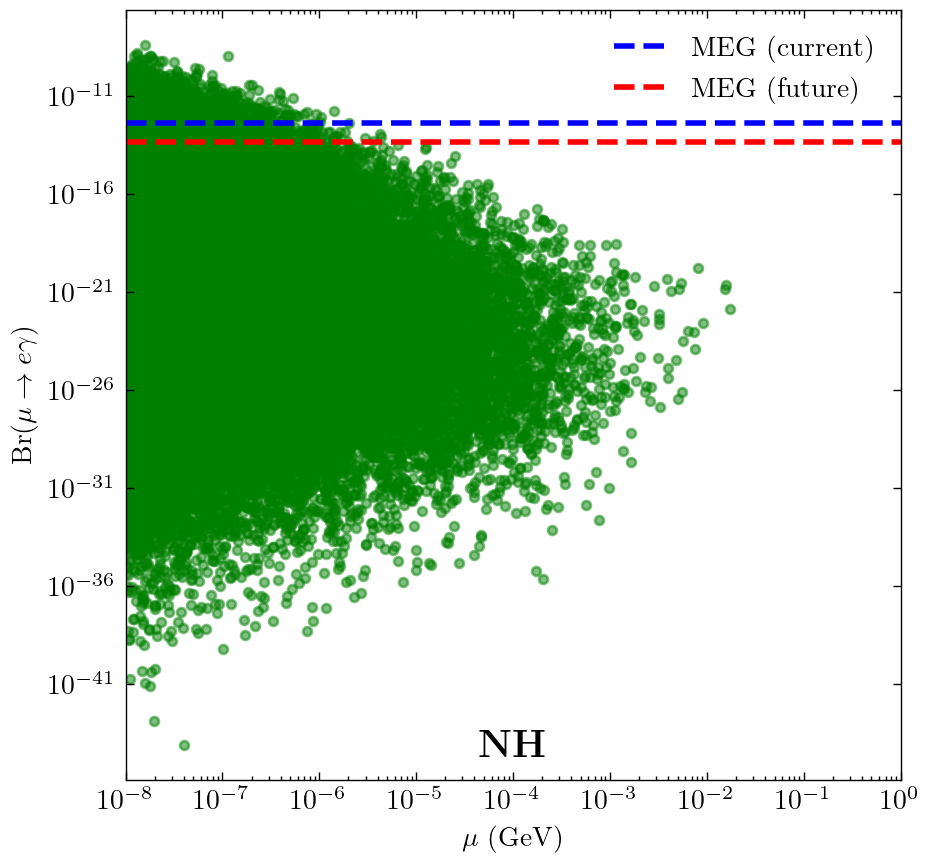}
\caption{The branching ratio of $\mu \rightarrow e \gamma$ as a function of the ESS parameter
$\mu$ for active neutrinos in the normal hierarchy (NH). The red and the blue dashed lines represent the MEG II experiment current bound ${\rm BR}(\mu \to e \gamma) < 3.1 \times 10^{-13}$~\cite{MEGII:2023ltw} and future bound $\text{BR} (\mu \to e \gamma) < 6 \times 10^{-14}$~\cite{MEGII:2018kmf}, respectively. 
}
\label{fig:LFV2}
\end{figure}

In  Fig.~\ref{fig:LFV2}, we show the $\mu\to e\gamma$ branching ratio  
as a function of $\mu$ for the active neutrinos in the normal hierarchy (NH). The red and blue dashed lines represent the current and future upper bounds on $\text{Br}(\mu \rightarrow e \gamma)$ given by the MEG II experiment~\cite{MEGII:2023ltw}. 
From Eq.~\eqref{eq:g-factor}, $\text{Br}(\mu \rightarrow e \gamma)$ is proportional to $\left| G_{\mu e} \right|^2$. As discussed in Sec.~\ref{sec:constraints}, $\Theta \sim (\widehat M_\nu/\mu)^{1/2}$, it follows that $\left| G_{\mu e}\right| \propto 1/\mu$. Therefore, the sensitivity to $\mu$ is increased by a factor of $\sim2.3$ given the anticipated experimental sensitivity.
From Fig.~\ref{fig:LFV2}, we find that the current cLFV searches are sensitive to $\mu \lesssim 2$~keV, which is slightly improved to $\mu \lesssim 5$~keV by the upcoming  MEG II experiment.
Similar conclusions are drawn in the inverted hierarchy (IH), the result of which is presented in Appendix~\ref{app:paramscanIH}. 

As mentioned in Sec.~\ref{sec:constraints}, current searches for $\mu \to eee$ and $\mu -e$ conversion cannot compete with $\mu \to e\gamma$. The planned searches for $\mu \to eee$ and $\mu -e$ conversion with ultimate future sensitivities are expected to give stronger constraints on $|G_{\mu e}|$  by a factor of $\sim 1.9$ and 6, respectively, given the relations in Eq.~\eqref{eq:correlation}.
Therefore, we can estimate that the planned searches for $\mu \to eee$ and $\mu -e$ conversion are sensitive to the regions $\mu \lesssim 10$~keV and $\mu \lesssim 30$~keV, respectively.

\begin{figure}[htb!]
\centering
\includegraphics[width=0.8\linewidth]{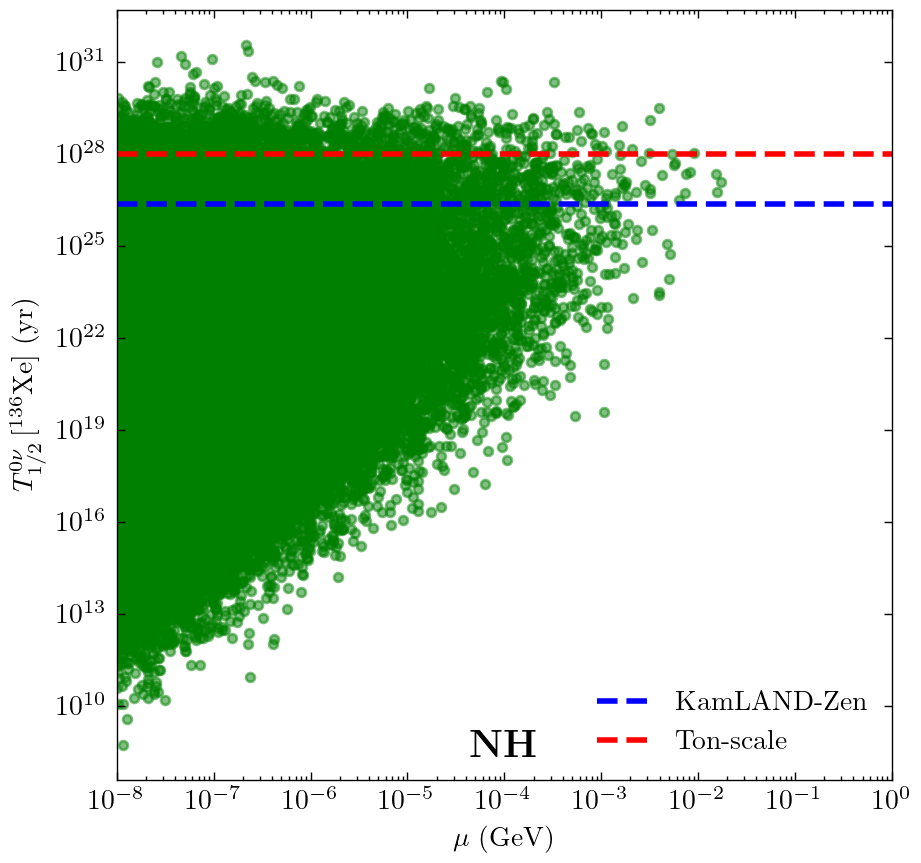}
\caption{The $\onbb$ decay half-life for $^{136}{\rm Xe}$ as a function of $\mu$ for active neutrinos in the normal hierarchy (NH). The red and blue dashed lines represent the limits set by the KamLAND-Zen experiment and future ton-scale experiments for $T_{1/2}^{0\nu} >2.3\times 10^{26}$ yrs and $10^{28}$~yrs, respectively.
}
\label{fig:onbbscan}
\end{figure}

In contrast to the situation for cLFV searches, the region of $\mu\gtrsim 5$~keV can be probed by $\onbb$ decay. In Fig.~\ref{fig:onbbscan}, we present the predicted inverse $\onbb$ decay half-life for $^{136}$Xe as a function of $\mu$ in the NH.
The red and blue lines represent, 
respectively, the current KamLAND-Zen~\cite{KamLAND-Zen:2022tow} and future ton-scale experiments' lower limits~\cite{Han:2017fol,Gomez-Cadenas:2019sfa,LEGEND:2017cdu,nEXO:2017nam}.
We find that the predicted $\onbb$ decay within a large portion of parameter space, where $\mu \lesssim 10~{\rm MeV}$, is in the reach of ton-scale $\onbb$-decay experimental searches.

In the IH, however, the theoretical predictions differ significantly. As illustrated in Fig.~\ref{fig:onbbscan_IH} in Appendix~\ref{app:paramscanIH}, the $\onbb$ decay rates are notably larger than those in the NH. This situation is analogous to the scenario of $\onbb$ decay in the standard mechanism~\cite{Agostini:2022zub}. Consequently, future ton-scale $\onbb$-decay experiments can probe nearly the entire parameter space for $\mu \lesssim 10\mev$.

So far, we have illustrated the sensitivities of $\onbb$ decay and cLFV searches using the model parameter $\mu$. As indicated in Eqs.~\eqref{eq:MSbd}~\eqref{eq:CI_MS}, if $\mu$ is within the keV-MeV range, the sterile neutrinos could be at MeV scale, 
These light sterile neutrinos give significant contributions to $\onbb$ decay as depicted in Figs.~\ref{fig:onbbscan}~and~\ref{fig:onbbscan_IH}. On the other hand, 
there coexists heavy sterile neutrino with a mass of around 100~GeV or even heavier, whose contribution dominates over the others to the decay branching ratio of $\mu\to e\gamma$, as discussed below Eq.~\eqref{eq:loopfactor}.

\subsection{Low-scale leptogenesis}
We study the viability of low-scale leptogenesis using the parameter ranges in Tab.~\ref{tab:param_scan}. Notice that, after diagonalizing the full neutral-fermion mass matrix, an effective Yukawa interaction arises between the sterile states $S$, the SM Higgs boson, and the lepton sector. At leading order, the effective coupling of this interaction is given by
\begin{equation}
    Y^{\rm eff}_S \simeq Y_{D}\,M_S\,M_R^{-1}\,.
\end{equation}
To prevent these interactions from thermalizing $S$ during the leptogenesis epoch, we require the corresponding reaction rates (scattering and inverse decays involving $S$) to remain below the Hubble expansion rate, leading to the stringent bound $Y^{\rm eff}_S \lesssim 10^{-7}$~\cite{Laine:2022pgk}, which translates into an equilibration temperature
\begin{equation}
    T_{\rm eq} \sim \frac{(Y^{\rm eff}_S)^2\, M_{\rm Pl}}{1.66\,\sqrt{g_*}} < T_{\rm EW}\sim\mathcal{O}(100\,{\rm GeV})\,,
\end{equation}
where $M_{\rm Pl}=1.22\times10^{19}\,{\rm GeV}$ is the Planck mass and $g_*=g_{SM}= 106.75$ is the total number of degrees of freedom.
Accordingly, sterile-mediated washout processes remain out of equilibrium throughout the electroweak epoch, ensuring that the generated lepton asymmetry is preserved.

\begin{figure}[htb!]
\centering
\includegraphics[scale=0.5]{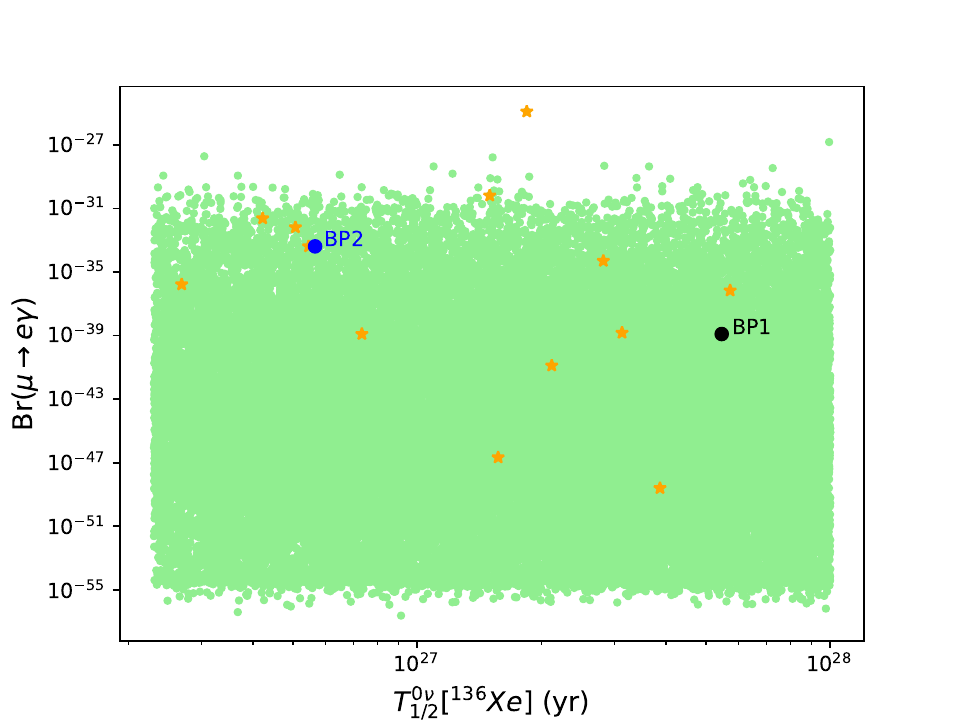}
\caption{Allowed parameter space from the model prediction of the branching ratio of $\mu \rightarrow e \gamma$ and $\onbb$ decay. While the light green points indicate the parameter space, which is consistent with ton-scale experimental sensitivities of $\onbb$-decay but are beyond the reach of the MEG II cLFV experiment, the orange points represent which can give the correct BAU along with the other constraints.}
\label{fig:Combined_NH}
\end{figure}

We compute the baryon asymmetry $Y_B$ as a function of the ESS parameter $\mu$ by numerically solving the Boltzmann equations described above and marginalizing over the remaining model parameters within the ranges specified earlier. For the normal hierarchy (NH), we find that reproducing the observed baryon asymmetry, $\eta^{\text{obs}} = (6.12 \pm 0.04) \times 10^{-10}$~\cite{ParticleDataGroup:2020ssz, Planck:2018vyg}, requires $\mu \lesssim \mathcal{O}(\text{MeV})$, with viable parameter points also lying within the sensitivity of current and forthcoming charged-lepton-flavor-violation and $\onbb$ decay searches. We explore in further detail the interplay between $\onbb$-decay, cLFV, and low-scale leptogenesis in the NH scenario as shown in Fig.~\ref{fig:Combined_NH}. A similar plot in the inverted hierarchy (IH) scenario is provided in Fig.~\ref{fig:Combined_IH} in Appendix~\ref{app:paramscanIH}. Points shown in light green lie within the reach of ton-scale $\onbb$-decay experiments but remain beyond the projected sensitivity of MEG~II. From this set, we identify parameter points that reproduce the observed baryon asymmetry within $\pm 10\%$ and highlight them in orange. These regions demonstrate that the ESS framework can simultaneously accommodate successful low-scale leptogenesis and observable $\onbb$-decay. 

To make the discussion more quantitative, we select two of the orange points that predict a baryon asymmetry $Y_B$ exceeding the observed value, $Y_B^{\mathrm{obs}}$, and adjust the relevant parameters to reproduce $Y_B^{\mathrm{obs}}$. The resulting benchmark points, denoted BP1 and BP2, are displayed in Fig.~\ref{fig:Combined_NH} as black and blue dots, respectively. Their corresponding parameter values are listed in Tab.~\ref{tab:bps}. We note that both BP1 and BP2 satisfy the direct-search constraint
$\mu\gtrsim 0.5\,\mathrm{keV}$ discussed in Sec.~\ref{sec:bbn_&_direct}.

\begin{table}[hbt!]
\centering
\begin{tabular}{lrr}
\toprule
\textbf{Parameter} & \textbf{BP1} & \textbf{BP2}\\
\midrule
$a_i$ & $0.058$ & $3.06$ \\
$\phi_{i}$ & $0.058$ & $3.06$ \\
$\lambda_1$, $\lambda_2$ [rad] & $0$ & $0$\\ 
$m_{\nu {\rm min}}$ [eV] & $0.0129$ &  $0.000502$ \\
$\mu$ [keV] & $65.20$ & $3.47$ \\
$M_{R1}$ [TeV] & $23.488$ & $48.171$ \\
$ M_{R2}$ [TeV] & $23.489$ & $48.174$ \\
$M_{R3}$ [TeV] & $1.378 \times 10^{5}$ & $5.91 \times 10^{10}$ \\
$v_{\phi}$ [GeV] & $9102$ & $7767$ \\
\bottomrule
\end{tabular}
\caption{Benchmark values of the parameters for BP1 and BP2, where $m_{\nu {\rm min}} = m_1$ in the NH and $m_3$ in the IH.}
\label{tab:bps}
\end{table}

As we have discussed earlier, our diagonalization procedure requires the Dirac mass matrix $M_S$ to be invertible, which implies that $M_D$ must also be invertible according to the extended Casas-Ibarra parametrization in Eq.~\eqref{eq:CI_YS}. Consequently, even though $N_{R1,2}$ are primarily responsible for the CP asymmetry, we cannot completely decouple $N_{R3}$.
For the reader's benefit, who may wish to reproduce our results, we have listed the values of $Y_D$ in Eq.~\eqref{eq:YDtextures}.
\begin{figure*}[htb]
\begin{align}
& (Y_D)_{\mathrm{BP1}} =
\begin{pmatrix}
2.45\times10^{-6} e^{-0.551 i} & 1.478\times10^{-2} e^{-2.028 i} & 7.415 \times 10^{-10} e^{0.706 i} \\
5.319 \times 10^{-8} e^{-0.104 i} & 4.633 \times 10^{-4} e^{0.852 i} & 1.250 \times 10^{-9} e^{-0.032 i} \\
8.37 \times 10^{-12} e^{0.415 i} & 5.159 \times 10^{-8} e^{-1.036 i} & 5.167 \times 10^{-11} e^{-1.111 i}
\end{pmatrix}, \nonumber \\
& (Y_D)_{\mathrm{BP2}} =
\begin{pmatrix}
6.57 \times 10^{-6} e^{2.052 i} & 6.722 \times 10^{-9} e^{0.338 i} & 1.557 \times 10^{-11} e^{-0.013 i} \\
9.64 \times 10^{-6} e^{0.037 i} & 2.977 \times 10^{-4} e^{0.193 i} & 5.683 \times 10^{-9} e^{2.716 i} \\
1.98 \times 10^{-9} e^{-2.301 i} & 4.70 \times 10^{-6} e^{2.394 i} & 4.232 \times 10^{-10} e^{2.290 i}
\end{pmatrix}.
\label{eq:YDtextures}
\end{align}
\end{figure*}
These textures are utilized to facilitate the parameter exploration for these benchmark points. With these parameter choices, we obtain $ (Y_B)^{\rm{BP1}} = 8.63 \times 10^{-11} $ and $ (Y_B)^{\text{BP2}} = 8.55 \times 10^{-11}$. Importantly, BP1 and BP2 also lie within the reach of future ton-scale $\onbb$ decay. 
The sterile neutrino mass scales and the corresponding active–sterile mixing parameters for the benchmark points are summarized in Appendix \ref{app:EstBPs}. Overall, both benchmark points yield sterile neutrinos with a broad mass hierarchy spanning sub-MeV to multi-TeV scales and the active-sterile mixings are consistent with low-scale seesaw expectations. Thus, these choices for BP1 and BP2 demonstrate the feasibility of leptogenesis in the presence of a next-generation $0 \nu \beta \beta$-decay observation. In principle, one could undertake a similar exercise for the other orange points in  Fig.~\ref{fig:Combined_NH}, varying the parameters modestly in order to obtain the precise value of $Y_B^\mathrm{obs}$.  In practice, doing so is costly in terms of time and computing resources, so we defer this exercise to future work.

\section{Conclusion}
\label{sec:conclusion}

In this work, we have considered the extended seesaw model where both right-handed neutrinos $N_R$ and left-handed sterile neutrinos $S_L$ are introduced, along with two mass scales of lepton number violation, i.e., the Majorana masses $M_R$ and $\mu$, and two Dirac masses $M_D$ and $M_S$. Under the assumptions of the mass hierarchies in Eq.~\eqref{eq:hier1}, one can diagonalize the neutrino mass matrix and obtain the active neutrino masses and the sterile neutrino masses, the latter of which are suppressed by $M_S M_R^{-1}$. 

We have demonstrated that this model can accommodate both viable low-scale leptogenesis and observable $\onbb$ decay, in contrast to the expectations based on generic arguments of the studies. The key ingredients for this exception to the EFT arguments are (i) the introduction of the ESS parameter $\mu$ that governs the overall scale of light neutrino masses and allows for consistency with the scale implied by experiment even with $\mathcal{O}(1)$ Dirac Yukawa couplings $(Y_D)$; (ii) 
suppression of the $N_R$ contribution to the leptogenesis washout rate by the scale of the $N_R$ masses, due either to the Boltzmann factor associated with on-shell external $N_R$ or the off-shell $N_R$ propagator after performing real intermediate state subtraction to the corresponding amplitudes; (iii) suppression of the washout rate contribution from the singlet scalar by a combination of the heavy $N_R$ masses, the VEV of $\phi$, and the values of elements of the Casas-Ibarra matrix $\mathcal{R}$; (iv) enhanced light sterile neutrino contributions to the $\onbb$ decay rate due to the ratio of sterile and light neutrino masses.  

In particular, the ratio of contributions to the washout rate goes as
\begin{align}
    \frac{\widetilde{W}}{W}&\sim \left(\frac{v_H}{v_\phi}\right)^2  \left(\frac{\mu}{\widehat{M}_\nu}\right)  \left(Y_D \mathcal{R}\right)^2 \nn\\
    &\qquad \times \left(\frac{m}{T}\right)^4 \left[\exp\left(-m/T\right)\right]^2  \;,
\end{align}
where $T\sim M_{R1}$, and $m$ denotes the larger one of $\widehat M_S$ and $M_\phi$. The standard and new contributions $W$ and $\widetilde W$ to the washout rate are defined in Eq.~\eqref{eq:washout} and Eq.~\eqref{eq:New_washout}, respectively.

On the other hand, Eqs.~(\ref{eq:mixing}, \ref{eq:CI_MS}, \ref{eq:WC}, \ref{eq:subamplitude}) imply that 
the ratio of sterile and light neutrino $\onbb$-decay amplitudes goes as 
\begin{equation}
\label{eq:DBDratio}
\frac{\mathcal{A}_L(m_{4,5,6})}{\mathcal{A}_L(m_{1,2,3})}\sim \left(\frac{{\widehat{M}}_\nu}{\mu}\right)\ \left( \frac{m_{4,5,6}}{m_{1,2,3}} \right) \ \ \ .
\end{equation}
To remind the reader, ${\widehat M}_\nu$ denotes elements of the diagonal, light (active) neutrino mass matrix.
While $\mu/{\widehat M}_\nu$ can vary from $[10^2, 10^8]$, the factors of $(v_H/v_\phi)^2$, $\mathcal{R}^2$, $(m/T)^4$ and $\left[\exp\left(-m/T\right)\right]^2$ can compensate to keep the ratio ${\widetilde{W}}/{W}$ sufficiently small.

On the other hand, the ratio $m_{4,5,6}/m_{1,2,3}$ can compensate for the ${\widehat M}_\nu/\mu$ suppression to allow for an observable impact of the light sterile neutrinos on the $\onbb$-decay rate. In effect, the factors  of ${\widehat M}_\nu$ and $m_{1,2,3}$ in Eq.~(\ref{eq:DBDratio}) cancel, leaving only a dependence on $m_{4,5,6}/\mu$.

In principle, the sterile neutrinos would contribute to low-energy observables such as the rates of charged lepton flavor violating (cLFV) decay $\mu \to e\gamma$ and $\onbb$ decay through the active-sterile mixing. In practice, after accounting for the experimental constraints on active-sterile mixing, we find that the parameter choices compatible with viable leptogenesis and potentially observable $\onbb$ decay signals in future ton-scale experiments generally predict cLFV rates below the sensitivity of current and planned searches.

\vskip1cm
{\large \bf Acknowledgements} 
\\[3mm]
GL is supported by the National Natural Science Foundation of China under Grant No. 12347105, the Guangdong Basic and Applied Basic Research Foundation (2024A1515012668), and the Fundamental Research Funds for the Central Universities, Sun Yat-sen University (23qnpy62), and SYSU startup funding. 
The work of MJRM and SS is supported in part under the US Department of Energy contract DE-SC0011095. SUQ acknowledges the support by the U.S. DOE under Grant No. DE-FG02-00ER41132. SS would like to thank Chayan Majumdar for helpful discussion and valuable inputs.


\appendix
\section{CP-violation and Boltzmann Equations}
\label{app:leptogenesis}
The general expression for the CP-asymmetry parameter $\epsilon_1$ considering all the contributions in Fig.~\ref{fig:LGdecay} is given by
\begin{align}
\epsilon_1 = \frac{1}{8\pi} \sum_{k \neq 1} & \Biggl\{\Big(g_V(x_k) + g_S(x_k)\Big) \frac{\text{Im}\left[(Y_D^\dagger Y_D)^2_{k1}\right]}{(Y_D^\dagger Y_D + Y_S^\dagger Y_S)_{11}} \nonumber \\
& + g_S(x_k) \frac{\text{Im}\left[(Y_D^\dagger Y_D + Y_S^\dagger Y_S)_{1k}\right]}{(Y_D^\dagger Y_D + Y_S^\dagger Y_S)_{11}}\Biggr\},
\label{eq:eps_full}
\end{align}

where
\begin{align}
g_V(x) &= \sqrt{x} \left\{1-(1+x)\ln \left( \frac{1+x}{x}\right)\right\}\,,\\
g_S(x) &= \frac{\sqrt{x}}{(1-x)}\,.
\end{align}
Here, $x_k = M_{Rk}^2/M_{R1}^2$ for $k \neq 1$. The second term comes from the interference of the tree-level diagram with diagram (d) in Fig.~\ref{fig:LGdecay}. When $x\gg1$, the self-energy diagrams become negligibly small, leading to the standard thermal leptogenesis scenario~\cite{Davidson:2002qv} where the asymmetry is associated with hierarchical right-handed neutrino masses~\cite{Chen:2007fv}. In contrast, when $x \simeq 1$, the contribution of the vertex to $\epsilon_1$ is considerably smaller compared to the contribution of the self-energy diagrams, leading to the resonant enhancement of the asymmetry in Eq.~\eqref{eq:eps_res}.

In general, the Boltzmann equations for $N_{R1}$ and the $(B-L)$ asymmetry can be formulated as follows~\cite{Plumacher:1996kc}
\begin{equation}
\frac{\text{d} Y_{N_{R1}}}{\text{d}z} = - \frac{z}{s H (M_{R1})} \Big[ \Big(\frac{Y_{N_{R1}}} {Y^{(eq)}_{N_{R1}}} - 1\Big)(\gamma_{D}+ 2 \gamma_{H,s} + 4 \gamma_{H,t})\Big],
\label{eq:BE1}
\end{equation}
\begin{align}
\frac{\text{d} Y_{B-L}}{\text{d} z}  & = - \frac{z}{s H (M_{R1})} \Biggl[ \left( \frac{1}{2} \frac{Y_{B-L}}{Y^{(eq)}_{\ell}} + \epsilon_1 \Big(\frac{Y_{N_{R1}}}{Y^{(eq)}_{N_{R1}}} - 1 \Big) \right) \gamma_{D}  \nonumber \\
& +\frac{Y_{B-L}}{Y^{(eq)}_{\ell}}\left\{2 \gamma_{N,s} + 2 \gamma_{N,t}\right\} \nonumber \\
& + \frac{Y_{B-L}}{Y^{(eq)}_{\ell}}  \left( 2 \gamma_{H,t} + \frac{Y_{N_{R1}}}{Y^{(eq)}_{N_{R1}}}  \gamma_{H,s} \right) \Biggr]
\end{align}
where $Y_{N_{R1}}^{(eq)}$ and $Y_{\ell}^{(eq)}$ are the $N_{R1}$ and lepton comoving densities at equilibrium, $z = {M_{R1}}/{T}$ and $H (M_{R1})$ is the Hubble parameter at $T = M_{R1}$ given by,
\begin{equation}
    H (T = M_{R1}) = 1.66 \sqrt{g_{\text{eff}}} \left( \frac{T^{2}}{M_{\text{Pl}}} \right)_{T = M_{R1}},
\end{equation}
with $M_{\text{Pl}}=1.22\times10^{19}\,{\rm GeV}$ being the Planck mass and $g_{\rm eff} = 106.75$ in the SM. The reaction densities $\gamma_D$, $\gamma_N$, and $\gamma_H$ consider the LNV scattering diagrams involving the lighter right-handed neutrino that are shown in Figs.~\ref{fig:LGscattering1}~and~\ref{fig:LGscattering2}. The full expressions can be found in Refs.~\cite{Giudice:2003jh, Granelli:2020pim, Harz:2021psp}.

\section{Washout Contributions}
\label{app:washout}
In the context of the washout scenario, the standard washout parameter can be expressed as
\begin{equation}
     K = \frac{\Gamma_{N_{R1}}}{H(z=1)},
\end{equation}
where the total washout in this model is given by
\begin{equation}
    W^{\rm{tot}}(z) = W(z, Y_{N_{R1}}) + \widetilde{W}(z),
\end{equation}
with the term $W(z, Y_{N_{R1}})$ representing the standard contribution and $\widetilde{W}(z)$ accounting for the additional contributions arising in this scenario. The standard contribution is described by
\begin{equation}
\label{eq:washout}
     W(z, Y_{N_{R1}}) = 2 [ N_{R1}(z) + S_t(z) ] \frac{Y_{N_{R1}}^{\rm{eq}}}{Y_{B-L}^{\rm{eq}}} + S_s(z) \frac{Y_{N_{R1}}}{Y_{B-L}^{\rm{eq}}},
\end{equation}
where $S_{s/t}(z)$ are the $\Delta L = 1$ scattering rates, and $N_{R1}(z) = N_s(z) + N_t(z)$ corresponds to the $\Delta L = 2$ scattering rates, given by
\begin{align}
& S_{s/t}(z) = \frac{\gamma_{S_{s/t}}}{zH(z)n_{N_{R1}}^{\rm{eq}}} = \frac{1}{zH(z)} \frac{M_{R1}}{48 \pi^2 \zeta(3) K_2(z)} I_{S_{s/t}}(z) \,, \\
& N_{R1}(z) = \frac{\gamma_{N_{R1}}}{zH(z)n_{N_{R1}}^{\rm{(eq)}}} = \dfrac{1}{zH(z)} \frac{M_{R1}}{48 \pi^2 \zeta(3) K_2(z)} I_{N_{R1}}(z),
\end{align}
with $I_{\{S_{s/t}, N_{R1} \}}(z) = \int_{x_{\min}}^\infty dx\ \sqrt{x}\,\hat \sigma_{\{S_{s/t}, N_{R1} \}}(x)\,K_1(\sqrt{x}\,z)\,$
%
and the reduced cross section is defined as $\hat\sigma \equiv 2s \lambda[1,m_i^2/s,m_j^2/s] \sigma$. Here, $\sigma$ represents the total cross section summed over initial and final states, $\lambda[a,b,c]$ is the Källén function, and $s$ denotes the square of the center-of-mass energy.

Moreover, the new contributions to the washout can be written as
\begin{equation}
    \widetilde{W}(z) = 2 \widetilde{N}_{s}(z) \frac{Y_{N_{R1}}^{\rm{eq}}}{Y_{B-L}^{\rm{eq}}},
    \label{eq:New_washout}
\end{equation}
with 
\begin{align}
\widetilde{N}_{s}(z) = \frac{\gamma_{\widetilde{N}_{R1}}}{zH(z)n_{N_{R1}}^{\rm{(eq)}}} = \frac{1}{zH(z)} \frac{M_{R1}}{48 \pi^2 \zeta(3) K_2(z)} I_{\widetilde{N}_{R1}}(z),
\end{align}
where $I_{\widetilde{N}_{R1}}(z) = \int_{x_{\min}}^\infty  dx \sqrt{x} \hat \sigma_{\widetilde{N}_{R1}}(x) K_1(\sqrt{x} z)$
and the scattering cross-section for this additional contribution can be written as
\begin{align}
   \sigma \equiv \frac{|Y_S|^2 |Y_D|^2}{32 \pi M_{R1}^2} \Bigg[ \frac{(s - {\widehat M_S}^2 - M_\phi^2 )^2}{(s- (\widehat M_S + M_\phi)^2) (s- (\widehat M_S - M_\phi)^2)} \Bigg]^{1/2}.
   \label{eq:new_cs}
\end{align}

\section{Parameter scan results in the inverted hierarchy}
\label{app:paramscanIH}

\begin{figure}[htb!]
\centering
\includegraphics[width=1.0\linewidth]{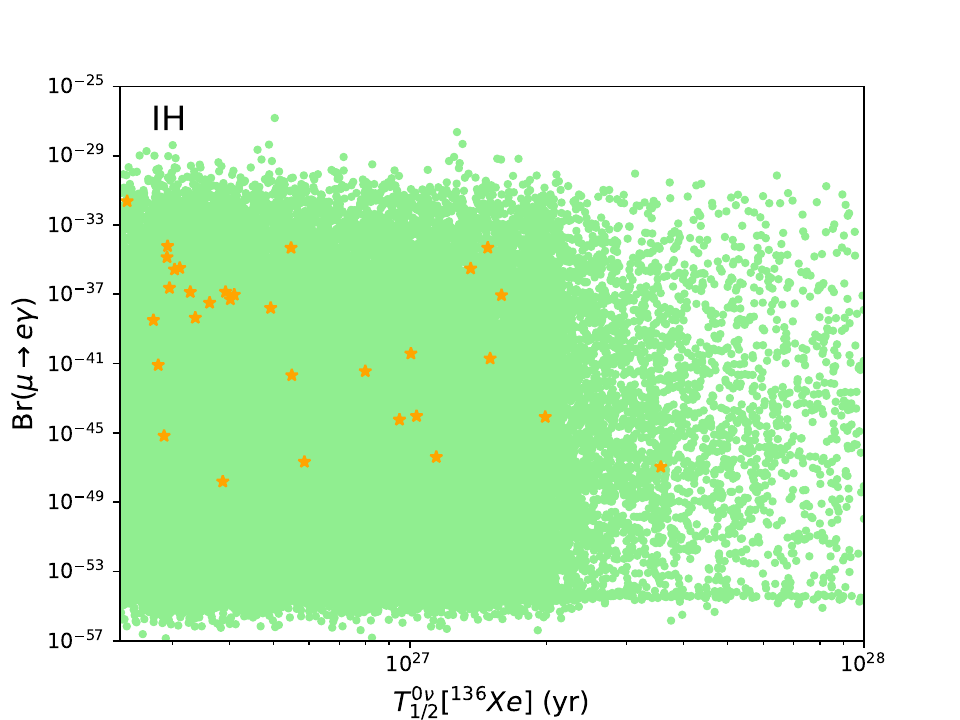}
\caption{Same as Fig.~\ref{fig:Combined_NH}, but for active neutrinos in the inverted hierarchy.}
\label{fig:Combined_IH}
\end{figure}

Here we present a version of Figs.~\ref{fig:LFV2},~\ref{fig:onbbscan},~and~\ref{fig:Combined_IH} for active neutrinos in the inverted hierarchy.

\begin{figure}[htb!]
\centering
\includegraphics[width=0.8\linewidth]{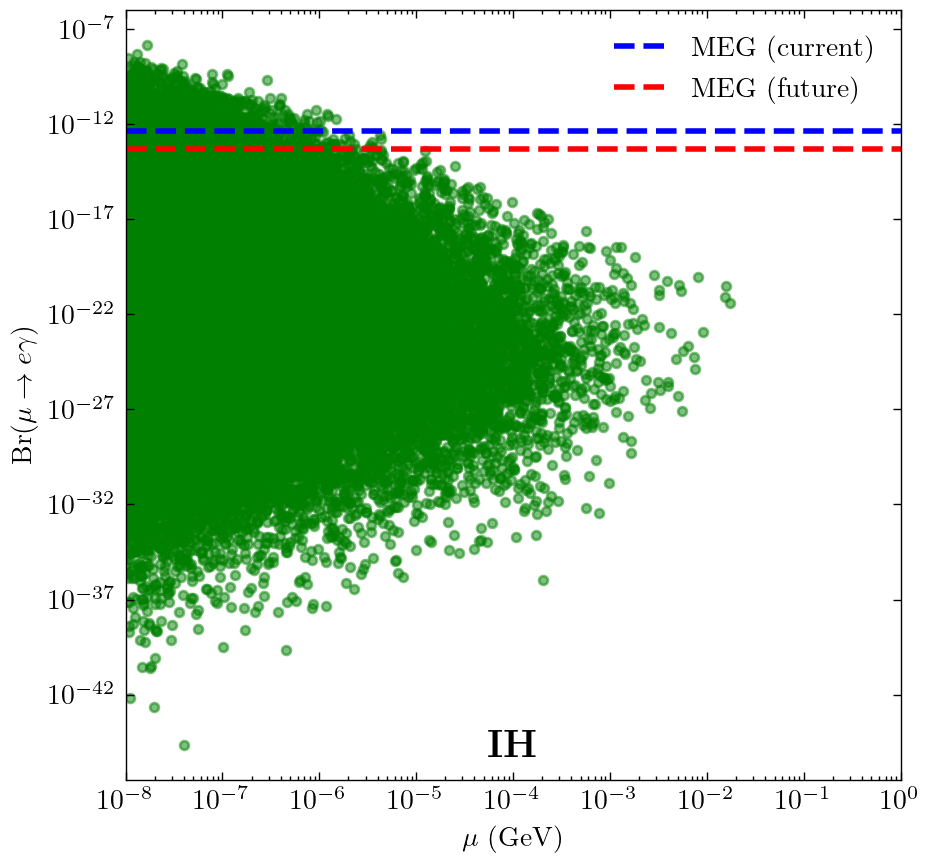}
\caption{Same as Fig.~\ref{fig:LFV2}, but for active neutrinos in the inverted hierarchy.
}
\label{fig:LFV2_IH}
\end{figure}

\begin{figure}[htb!]
\centering
\includegraphics[width=0.8\linewidth]{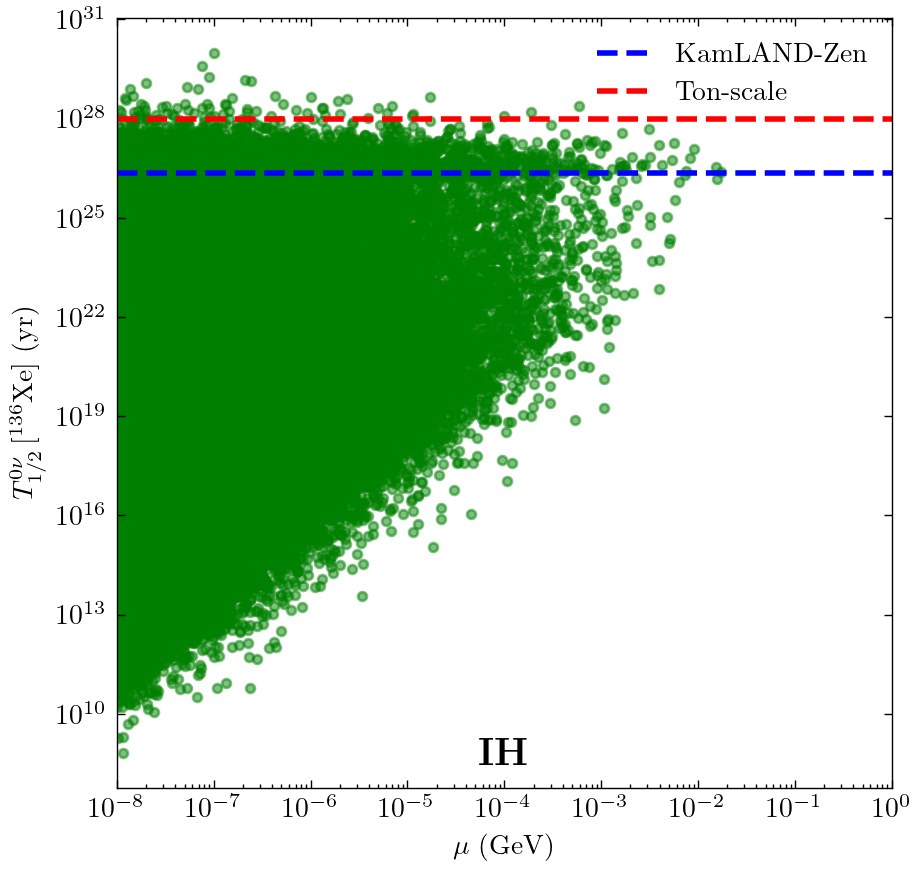}
\caption{
Same as Fig.~\ref{fig:onbbscan}, but for active neutrinos in the inverted hierarchy.
}
\label{fig:onbbscan_IH}
\end{figure}

\section{Sterile Neutrino Masses and Active–Sterile Mixing for BPs}
\label{app:EstBPs}
For completeness, we summarize the characteristic ranges of the heavy sterile neutrino mass matrices and the corresponding active–sterile mixing parameters for the two benchmark points used in our numerical analysis. For BP1, the entries of the heavy neutrino mass matrix span a wide range, $10^{-5}~ \text{GeV} \lesssim M_S \lesssim 10^{3}~ \text{GeV}$, with active-sterile mixing elements in the range $\Theta \sim (1.58-6.49) \times 10^{-4}$. In contrast, for BP2, the heavy neutrino mass matrix span, $10^{-2}~ \text{GeV} \lesssim M_S \lesssim 10^{2}~ \text{GeV}$, with comparatively large mixings of order $\Theta \sim \mathcal{O}(10^{-1})$ and the largest element is $\Theta_{\max} = 0.22$. The flavor-summed mixings, $U_{\alpha S}^2 = \sum_i |\Theta_{\alpha i}|^2$ are
\begin{align}
\text{BP1:}& \quad (U_{eS}^2, U_{\mu S}^2, U_{\tau S}^2)
\simeq (2.46,\,4.69,\,5.30)\times10^{-7}, \nonumber \\
\text{BP2:} & \quad (U_{eS}^2, U_{\mu S}^2, U_{\tau S}^2)
\simeq(0.032,\,0.149,\,0.047). \nonumber
\end{align}

\newpage

\bibliographystyle{apsrev4-1}
\bibliography{onbb_ESS}
\end{document}